\documentclass[12pt, draftclsnofoot, onecolumn]{IEEEtran}

\normalsize

\ifCLASSINFOpdf
\else
\fi
\usepackage{comment} 
\usepackage{lipsum} 
\usepackage{fullpage} 
\usepackage{mathtools}

\usepackage{cite}
\usepackage{amsmath,amssymb,amsfonts}
\usepackage{textcomp}
\usepackage{amsmath, bm}
\usepackage[pdftex]{graphicx}
\usepackage{algorithmic}
\usepackage{algorithm}
\usepackage{multicol}
\usepackage{graphicx}
\usepackage{lipsum, color}
\usepackage{mathtools, cuted}
\usepackage{commath}

\usepackage{epstopdf}
\usepackage{subfigure}
\usepackage[short]{optidef}
\usepackage[utf8]{inputenc}
\usepackage{mathtools, nccmath}
\usepackage{epsfig}
\DeclareGraphicsExtensions{.eps}
\usepackage{multirow} 

\usepackage{subfig}

\DeclareMathOperator*{\argmin}{argmin}

\usepackage{xcolor, soul}
\sethlcolor{red}

\usepackage [english]{babel}
\usepackage [autostyle, english = american]{csquotes}
\MakeOuterQuote{"}

\begin{document}
%
\title{Robust Hybrid Precoding for Beam Misalignment in Millimeter-Wave Communications}
%
%
%


\author{Chandan~Pradhan,~\IEEEmembership{Student Member,~IEEE,}
       Ang~Li,~\IEEEmembership{Member,~IEEE,}
       Li~Zhuo,~\IEEEmembership{Member,~IEEE,}
       Yonghui~Li,~\IEEEmembership{Fellow,~IEEE,}
       and~Branka~Vucetic,~\IEEEmembership{Fellow,~IEEE \vspace{-5.10ex}}
\thanks{Chandan Pradhan, Ang Li, Yonghui Li and Branka Vucetic are  with  the  Centre  of  Excellence  in  Telecommunications, School of Electrical and Information Engineering, University of Sydney, Sydney, NSW 2006, Australia (e-mail: chandan.pradhan@sydney.edu.au; ang.li2@sydney.edu.au; yonghui.li@sydney.edu.au; branka.vucetic@sydney.edu.au). Li Zhuo is with Beijing University of Technology, Beijing, China (email: zhuoli@bjut.edu.cn).}
}
\maketitle

\begin{abstract}

In this paper, we focus on the phenomenon of beam misalignment in Millimeter-wave (mmWave) multi-receiver communication systems, and propose robust hybrid precoding designs that alleviate the performance loss caused by this effect. We consider two distinct design methodologies:  I) the synthesis of a `flat mainlobe' beam model which maximizes the minimum effective array gain over the beam misalignment range, and II) the inclusion of the `error statistics' into the design, where the array response incorporating the distribution of the misalignment error is derived. For both design methodologies, we propose a hybrid precoding design that approximates the robust fully-digital precoder, which is obtained via alternating optimization based on the gradient projection (GP) method.  We also propose a low-complexity alternative to the GP algorithm based on the least square projection (LSP), and we further deploy a second-stage digital precoder to mitigate any residual inter-receiver interference after the hybrid analog-digital precoding. Numerical results show that the robust hybrid precoding designs can effectively alleviate the performance degradation incurred by beam misalignment.
\end{abstract}



%
\IEEEpeerreviewmaketitle

\section{Introduction}

\IEEEPARstart{T}he availability of rich spectrum in the millimeter-wave (mmWave) frequency bands makes mmWave communication one of the most promising candidates for future wireless communication systems to address the current challenge of bandwidth shortage \cite{pi2011introduction, rappaport2013millimeter, pi2016millimeter, alkhateeb2014channel, barati2016initial, Ren70Mis2015, cheng2018coverage,  rappaport2014millimeter}. Specifically,  the bands from 30 GHz to 300 GHz have been considered as the  primary contender for the future 5G network \cite{rappaport2014millimeter}. While mmWave signals are vulnerable to path loss, penetration loss and rain fading compared to the signals in the sub-6 GHz bands \cite{rappaport2013millimeter}, the short wavelength at mmWave frequencies allows large antenna arrays to be packed at the radio frequency (RF) front end for mmWave transceivers. The deployment of a large-scale antenna array enables the exploitation of highly directional beamforming to combat the  attenuation  from the environment \cite{pi2016millimeter, hur2013millimeter, alkhateeb2014channel, yu2016alternating}.

While it is possible to employ a fully-digital precoder in traditional sub-6GHz bands, it is unfortunately not promising to consider fully-digital processing for mmWave communications, due to the prohibitive cost and power consumption of the hardware components working at mmWave bands. To address this problem and implement mmWave communications both cost-efficiently and energy-efficiently, the concept of hybrid analog-digital structure has been introduced in  \cite{ alkhateeb2014channel}, which provides a promising trade-off between cost, complexity, and capacity of the mmWave network. The underlying principle behind the hybrid structure is to employ a reduced number of RF chains at the transceivers and divide the signal processing into an analog part and a digital part. Accordingly, the data streams at the mmWave transceivers are firstly processed by a low-dimension digital precoder, followed by the processing of a high-dimension analog precoder \cite{yu2016alternating, alkhateeb2015limited, sohrabi2016hybrid}. {For the analog precoding, low-cost phase shifters are commonly used \cite{hur2013millimeter}, which imposes a constant modulus requirement on the analog precoding matrix. Due to this constraint, the performance of hybrid precoder is usually inferior to the fully-digital precoder. In addition, analog precoding based on switches has also been considered in \cite{mendez2016hybrid, alkhateeb2016massive} as an alternative to constant modulus phase shifters.


There have been many recent works on hybrid precoding design in mmWave systems \cite{alluhaibi2017hybrid, yu2016alternating, yu2016alternatingLASSO, mendez2016hybrid, alkhateeb2016massive, alkhateeb2015limited, sohrabi2016hybrid, liang2014low, nguyen2016hybrid,angSVD2017,angVirtual2017751}. A commonality in these works is the attempt to maximize the overall spectral efficiency of the network with the assumption of perfect channel state information (CSI), which implicitly assumes perfect alignment between the transmitting and receiving beams. However, in practical mmWave scenarios where perfect CSI is usually not available \cite{el2014spatially, zhangTracking}, the estimation errors in the angle of arrival (AoA) or angle of departure (AoD) result in beam misalignment. In addition to the channel estimation errors, the imperfection in the antenna array, which includes array perturbation and mutual coupling \cite{yu2016alternating,yang2016analysis, li2006outage}, also contributes towards the imperfect alignment of beams.  Besides, environmental vibrations such as wind, moving vehicles, etc, can also be potential sources of beam misalignment \cite{hur2013millimeter}. The deployment of a large-scale antenna array that generates narrow beams for mmWave communications also makes the system  highly sensitive to beam misalignment.

To investigate the effect of imperfect alignment between the transmitting and receiving beams, existing studies in \cite{yang2016analysis, thornburg2015ergodic, cheng2018coverage, Ren70Mis2015} focus on analyzing the performance loss in terms of  ergodic capacity. The works in \cite{cheng2018coverage, Ren70Mis2015} evaluate the coverage performance of mmWave cellular networks with imperfect beam alignment, where \cite{cheng2018coverage} adopts an enhanced antenna model  that is able to express the mainlobe beamwidth and array gain as a function of the number of antennas. With the 3GPP two-dimension directional antenna model, the impact of beam misalignment on the performance of a 60GHz wireless network was studied in \cite{yang2016analysis}, where the probability distribution of the signal-to-interference-plus-noise ratio (SINR) was derived. For a mmWave ad-hoc network, the authors in \cite{thornburg2015ergodic} have derived a closed-form  expression for the ergodic capacity per receiver to quantify the performance loss due to the alignment error between the transmitting and receiving beams.

While there are already works that investigate the performance loss of beam misalignment, there are only a limited number of studies that consider the robust hybrid precoding design in the presence of beam misalignment \cite{hur2013millimeter,Robust2018Precoding}. In \cite{hur2013millimeter}, the authors consider the beam misalignment for backhaul links in small-cell scenarios, and propose a beam alignment method based on adaptive subspace sampling and hierarchical beam codebooks.  Nevertheless, only single-user transmission with analog-only processing was considered. Furthermore, the frequent beam re-alignment for a large-scale antenna array may not be favorable for delay-sensitive applications in mmWave communications. In \cite{Robust2018Precoding}, a hybrid precoding scheme is proposed to resist the AoA estimation errors based on the null-space projection in the analog  domain and diagonal-loading method in the digital domain, respectively. However, this scheme is only applicable to a single-user  mmWave communication system with the partially-connected structure, where each RF chain is connected to a subset of antennas. For the robust design against beam misalignment, the analysis in  \cite{cheng2018coverage, yang2016analysis} and \cite{thornburg2015ergodic} has established that the ideal `flat mainlobe' \footnote{A  constant  large  antenna  gain  within  the  narrow  mainlobe  and  zero  elsewhere.} model is robust to the loss in array gain, especially in the case of extremely narrow beams.  Accordingly, the ideal `flat mainlobe' model is theoretically conducive to alleviate the loss caused by beam misalignment. While there has been a previous attempt to synthesize a realistic `flat mainlobe' beampattern in \cite{cao2017constant}, it considers only a single-receiver analog beamforming and the relaxation of its optimization problem does not guarantee element-wise constant modulus for the analog precoder, which is required for mmWave transceivers that employ phase shifters. In addition, the statistics of the AoD/AoA estimation errors has been studied in  \cite{thornburg2015ergodic} and \cite{yu2006performance}, and it is shown in \cite{Robust2018Precoding} that the inclusion of the `error statistics' into the precoding design can also lead to an improved performance for the case of beam misalignment. However, the above two concepts  have not been well explored for robust multi-receiver hybrid precoding design in mmWave communications.

Motivated by this, in this paper we propose robust hybrid precoding schemes for a generic multi-receiver mmWave communication system. We consider the hybrid precoding design that approximates the robust fully-digital precoder by minimizing their Euclidean distance. To suppress the residual inter-receiver interference, we further introduce a second-stage digital precoder based on zero-forcing. For the precoding design, we consider two distinct methodologies to incorporate the beam misalignment error: a) the robust design based on the `flat mainlobe' model, and b) the robust design based on the prior knowledge of the `error statistics' in beam alignment.

The main contributions of the paper are summarized as follows:

\begin{itemize}

\item We develop two robust fully-digital precoders (DPs) based on the `flat mainlobe' model and the  `error statistics' for a multi-receiver mmWave system to alleviate the performance loss owing to beam misalignment.  The DP based on `flat mainlobe' model aims to maximize the minimal array gain for the receivers over the expected beam misalignment range. The resulting max-min non-convex optimization is solved in two steps: we firstly formulate an equivalent min-max problem with  the zero-forcing principle that fully cancels the inter-receiver interference, which is further transformed into a second-order cone programming (SOCP). For the robust DP based on `error statistics', we analytically derive the expected array response for the transmitter incorporating the beam alignment error distribution, which we utilize to obtain the closed-form robust DP that maximizes the array gain subject to zero inter-receiver interference.

\item Based on the obtained robust fully-digital precoder, we introduce the hybrid precoding design by minimizing the Euclidean distance between the hybrid precoder and the fully-digital precoder. The resulting optimization  is decoupled into two sub-problems and solved via alternating optimization, where the digital precoder and the analog precoder are obtained using the least-square approximation and the gradient projection (GP) method, respectively. A low-complexity scheme based on least square projection (LSP) is also introduced with a closed-form expression of the analog precoder for further complexity reduction. 


\item  We further design a second-stage digital precoder to fully mitigate the residual inter-receiver interference that is incurred due to the approximation process involved in the precoding design. The second-stage digital precoder applies a channel inversion on the effective channel of each receiver.

\item Our complexity analysis for the analog precoding design reveals that the computational cost of the LSP method is significantly lower than the GP method, while the GP method requires only approximately $30\%$ to $40\%$ of the computational cost compared to the scheme based on manifold optimization (MO) proposed in \cite{yu2016alternating}. Moreover, numerical results show that significant performance gains can be observed for the proposed robust designs compared to the non-robust hybrid precoders in the presence of imperfect beam alignment.

\end{itemize}

The rest of the paper is organized as follows. Section II describes the system model, channel model  and beam alignment error model. In Section III, the robust fully-digital precoder design and its approximation by the hybrid precoding are presented. Section IV introduces the second-stage digital precoder that cancels the inter-receiver interference. Numerical results are presented in Section V, and we conclude the paper in Section VI.   

\textit{Notations}: Bold upper-case letters  $\mathbf{Y}$, bold lower-case letters $\mathbf{y}$ and letters $y$ denote matrices, vectors and scalars respectively; $Y_{i,j}$ is the entry on the $i$-th row and $j$-th column of $\mathbf{Y}$; Conjugate, transpose and conjugate transpose of $\mathbf{Y}$ are represented by $\mathbf{Y}^*$, $\mathbf{Y}^T$ and $\mathbf{Y}^H$; $\mathbf{\norm{Y}}_F$ denotes the Frobenius norm of $\mathbf{Y}$; $\mathbf{Y}^\dagger$ is the Moore-Penrose pseudo inverse of $\mathbf{Y}$; $blkdiag\{ {\bf Y}_1,...,{\bf Y}_n \}$ denotes a block diagonal matrix with matrices ${\bf Y}_i$ on the block-diagonal; $vec(\mathbf{Y})$ indicates vectorization; $\norm{\bf y}_2$ is the $l_2$ norm of the vector ${\bf y}$; Expectation of a complex variable is noted by $\mathbb{E}[\cdot]$; $\odot$ and $\otimes$ denote the Hadamard and Kronecker product of two matrices; $\mathbf{I}$ is the identity matrix; $|\cdot|$ returns the absolute value of a complex number; $\angle$ denotes the argument of a complex number; $\mathcal{R}$ denotes the real part of a complex number.

\section{System Model}


\begin{figure}
\centering
\includegraphics[scale=0.4]{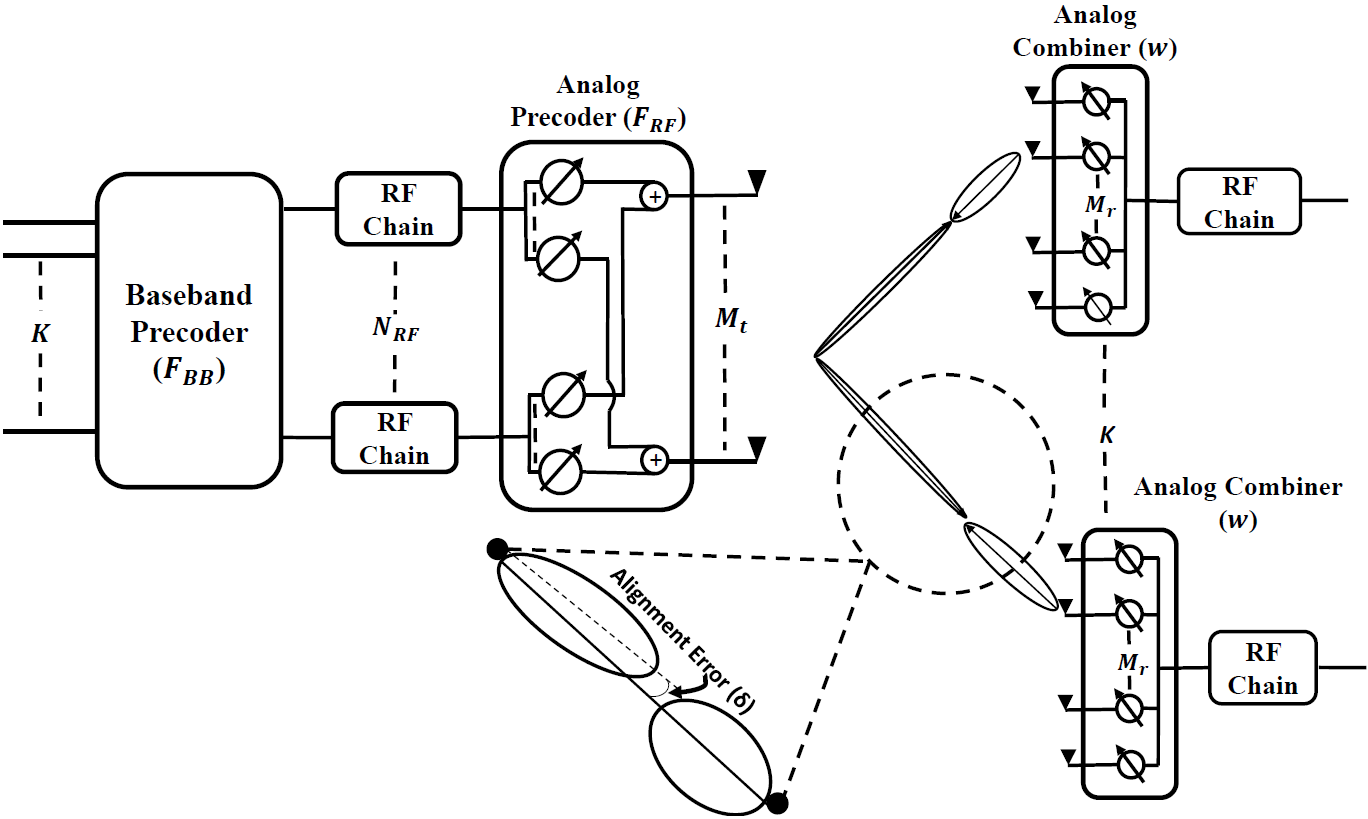}
\vspace{-0.5em}
  \caption{{ \small Block diagram for a multi-receiver mmWave MIMO system}}
\vspace*{-2mm}
\end{figure}

\subsection{System Model}

 We consider a multi-receiver mmWave system in the downlink, as shown in Fig.1 where a base station (BS) with $M_t$ antennas and $N_{RF}$ RF chains is communicating with $K$ receiver units (RUs), where $ K \leq N_{RF} \ll M_t$ is assumed at the BS to support simultaneous transmission with $K$ RUs. Each RU is equipped with $M_r$ antennas and a single RF chain, i.e., a single-stream transmission is assumed for each RU.} During transmission, the BS applies a $N_{RF} \times K$ digital precoder $\mathbf{F_{BB}}=\left[{\bf f}_1^{\bf BB}, {\bf f}_2^{\bf BB}, \dots,{\bf f}_K^{\bf BB} \right]$  followed by an $M_t \times N_{RF}$ analog precoder $\mathbf{F_{RF}} = \left[{\bf f}_1^{\bf RF}, {\bf f}_2^{\bf RF}, \dots, {\bf f}_{N_{RF}}^{\bf RF}\right]$, and the signal vector to be transmitted is therefore 

\begin{equation}
\begin{split}
\mathbf{x} &= \mathbf{F_{RF}F_{BB}s} = \mathbf{Fs},
\end{split}
\end{equation}
where $\mathbf{s} = \left[s_1, s_2, \dots, s_K \right]^T$, $s_k$ is the symbol transmitted to the $k$-th RU and $\mathbb{E}\left[{\bf s} {\bf s}^H \right] = \frac{P}{K} {\bf I}_K$. $P$ is the total transmit power at the BS, and in this work we have assumed uniform power allocation among different RUs. Since $\mathbf{F_{RF}}$ is implemented with analog phase shifters, its entries should satisfy the element-wise constant modulus constraint, i.e.,  $\left| \left[\mathbf{F_{RF}}\right]_{m,n} \right| = \sqrt{\frac{1}{M_t}}, \; \forall m,n$. The total power constraint is enforced by normalizing $\mathbf{F_{BB}}$ such that $\norm{\mathbf{F_{RF} F_{BB}}}^2_F = K$.



Based on the above, the received signal for the $k$-th RU is obtained as

\begin{equation}
\begin{split}
y_k &= \mathbf{w}_k^H {\bf H}_k \sum_{i=1}^{K} \mathbf{F_{RF}} {\bf f}_i^{\bf BB} s_i + {\bf w}_k^H {\bf n}_k, \\
\end{split}
\end{equation}
where ${\bf w}_k$ is the analog combiner for RU $k$, $\mathbf{H}_k$ is the $M_r \times M_t$ mmWave channel matrix between the BS and the $k$-th RU, and $\mathbf{n}_k \sim \mathcal{C N}(0, \sigma {\bf I})$ is the additive Gaussian noise at each RU. Similar to its  counterpart at the BS, each entry in $\mathbf{w}_k$ satisfies the constant modulus constraint.


\subsection{{Channel Model}}

MmWave channels are expected to be sparse with a limited number of propagation paths, and accordingly the channel between the BS and the $k$-th RU is given by \cite{alkhateeb2014channel}:

\begin{equation}
  \mathbf{H}_k = \sqrt{\frac{M_t M_r}{L}} \sum_{l = 1}^{L} {\gamma}_{k,l} {\bm \alpha} \left( \theta_{k,l}^{\left(AoA \right)} \right) {\bm \alpha} \left(\theta_{k,l}^{\left(AoD \right)} \right)^H, 
\end{equation}
where $L$ is the number of propagation paths between the BS and the $k$-th RU, and ${\gamma}_{k,l}$ is the complex gain of the $l$-th path following $\mathcal{CN}\left(0,\sigma^2_{\gamma}\right)$.  $\theta_{k,l}^{(AoD)}$ and $\theta_{k,l}^{(AoA)} \in [0, \pi]$ are the AoD and AoA along the $l$-th path, respectively, with ${\bm \alpha}\left(\theta_{k,l}^{(AoD)}\right)$ and ${\bm \alpha}\left(\theta_{k,l}^{(AoA)} \right)$ being the corresponding antenna array response vectors of the BS and the $k$-th RU, respectively. For uniform linear arrays (ULAs) considered in this paper, ${\bm \alpha}\left(\theta \right)$ for an $M$-element antenna array is given by

\begin{equation}
 {\bm \alpha}(\theta) = \frac{1}{\sqrt{M}} \left[1, e^{j \frac{2 \pi}{\lambda} d \; \cos(\theta)}, \dots, e^{j \frac{2 \pi}{\lambda} d \; (M - 1)\; \cos(\theta)} \right]^T,
\end{equation}
where $d$ and $\lambda$ are the antenna spacing and signal wavelength, respectively. Since transmission and reception at the mmWave system are done through highly directional beams and with single stream per RU, when multiple paths are available, it is reasonable to steer the beam toward the strongest path \cite{xiaoNOMA2018, almasi2019}. Therefore, in this work, we adopt a single path channel model as in \cite{almasi2019}, where the  channel described in (3) is reduced to

\begin{equation}
  \mathbf{H}_k = \sqrt{{M_t M_r}}  {\gamma}_{k} {\bm \alpha} \left( \theta_k^{\left(AoA \right)} \right) {\bm \alpha} \left(\theta_k^{\left(AoD \right)} \right)^H, 
\end{equation}
where ${\gamma}_{k}$, $\theta_{k}^{(AoD)}$ and $\theta_{k}^{(AoA)}$ are the channel gain, AoD and AoA of the strongest path between the BS and the $k$-th RU, respectively.


%

\subsection{Error Model for Beam Misalignment}

We define the beam misalignment error in AoA/AoD as

\begin{equation}
\delta = \theta - \hat{\theta},
\end{equation}
where $\theta$ is the actual AoA/AoD and $\hat{\theta}$ is the estimated AoA/AoD. Following  \cite{Robust2018Precoding, yu2006performance}, the beam alignment error $\delta$ is characterized by a random variable following a uniform distribution, given by

\begin{equation}
f(\delta) = 
\begin{cases}
    \frac{1}{2\beta},& \text{if } -\beta \leq \delta \leq \beta \\
    0,              & \text{otherwise}
\end{cases}
\end{equation}
where $\beta = \sqrt{3} \Delta$ and $\Delta$ represents the standard deviation of the beam alignment error. We assume the random misalignment error $\delta$ is bounded as $0 \leq |\delta| \leq \vartheta$, where $\vartheta$ is the mainlobe beamwidth of the transceiver units. Beam deviation exceeding $\vartheta$ is treated as alignment failure, and  appropriate methods based on the work in \cite{abari2016millimeter} can be used for realignment. In this paper, as we focus on the precoding design at the BS side, we employ the analog combiner that maximizes the array gain at each RU, given by \cite{almasi2019}

\begin{equation}
{\bf w}_k = {\bm \alpha} \left(\hat{\theta}_k^{(AoA)} \right), \; \forall k.
\end{equation}

\section{Hybrid Precoding Design by Approximating the Fully-Digital Precoder}


In this section, we present the hybrid precoding design by minimizing the difference between the hybrid precoder and the optimal robust fully-digital precoder. For the robust fully-digital precoder, we consider designs based on both the `flat mainlobe' model and the `error statistics' in the beam alignment.

\subsection{Robust Fully-Digital Precoder Design based on `Flat Mainlobe' (DP-FM)}

To alleviate the loss in the array again resulting from beam misalignment, the `flat-mainlobe' model aims to design the robust fully-digital precoder that maximizes the minimal array gain for each RU over the expected range of misalignment. Let $\mathbf{F_{FM}} = \left[{\bf f}_1^{\bf FM}, {\bf f}_2^{\bf FM}, ..., {\bf f}_K^{\bf FM} \right]$ be the $M_t \times K$  fully-digital precoder matrix, where $\mathbf{f}^{\bf FM}_k$ is the precoding vector for the $k$-th RU, and then the array gain of the BS corresponding to the $k$-th RU is given by

\begin{equation}
\left|\mathbf{{\bm \alpha}}\left(\varphi_k \right)^H {\bf f}^{\bf FM}_k \right|, \; \varphi_k \in \mathbf{\Phi}_k,
\end{equation}
where $\mathbf{\Phi}_{k}$ is the set of angular range covering the expected misalignment $\delta$ such that $\mathbf{\Phi}_k$ contains $N$ samples that are uniformly distributed in the range of $\left(\hat{\theta}_k^{(AoD)} -\beta\right) \leq \varphi_k \leq \left(\hat{\theta}_k^{(AoD)} + \beta\right)$. Let $\mathbf{A}_k = \left[{\bm \alpha} \left(\varphi_{k,1} \right), {\bm \alpha}\left(\varphi_{k,2} \right), ..., {\bm \alpha} \left(\varphi_{k,N} \right) \right]^H$ be the $N \times M_t$ matrix containing the array response of the BS in the range of $\hat{\theta}_k^{(AoD)} + \left|\beta \right|$, and with the goal of constructing a `flat mainlobe' beampattern for each $K$ RUs, we propose to maximize the minimum array gain over $\mathbf{\Phi}_k, \; \forall k$ by optimizing the robust fully-digital precoder subject to zero inter-receiver interference and transmit power limit, formulated as

\begin{equation}
\begin{aligned}
& \mathcal{P}_1: && \underset{\mathbf{F_{FM}}}{\text{max}} \; \; \underset{k \in \mathcal{K}}{\text{min}} \; {\norm{ {\bf A}_k {\bf f}_k^{\bf FM}}}^2_2  \\
& \text{\it s.t}
& & {\rm C1}: {\norm{ {\bf A}_k {\bf f}_j^{\bf FM}}}^2_2  = 0, \; \forall k, \; j\neq k \\
&&& {\rm C2}: \norm{\mathbf{F_{FM}}}_F^2 \leq K,
\end{aligned}
\end{equation}
where $\mathcal{K}$ is the set of $K$ RUs, ${\rm C1}$ cancels the inter-receiver interference and ${\rm C2}$ is the average transmit power constraint at the BS. The above problem can be equivalently reformulated into the following min-max form:

\begin{equation}
\begin{aligned}
& \mathcal{P}_2: && \underset{\mathbf{F_{FM}}}{\text{min}} \; \underset{k \in \mathcal{K}}{\text{max}} \; {\norm{ \left|{\bf A}_k {\bf f}_k^{\bf FM} \right| - {\bf d}(\varphi_k)}}^2_2 \\
& \text{\it s.t.}
& & {\rm C1}: {\norm{{\bf A}_k {\bf f}_j^{\bf FM}}}^2_2  = 0, \; \forall k, \; j\neq k \\
&&& {\rm C2}: \norm{\mathbf{F_{FM}}}_F^2 \leq K,
\end{aligned}
\end{equation}
where the vector ${{\bf d}(\varphi_k)} = \left[1, 1, \dots 1 \right]^T$ is to approximate a flat array gain over $\mathbf{\Phi}_k, \; \forall k$. The unit gain is attained when the beam is perfectly aligned in the direction of $\varphi_{k} \in \mathbf{\Phi}_k$, which is the upper bound for the array gain in the direction $\varphi_{k} \in \mathbf{\Phi}_k, \; \forall k$. Hence, $\mathcal{P}_1$ and $\mathcal{P}_2$ are `equivalent' in the sense that the minimization of $\mathcal{P}_2$ will maximize $\mathcal{P}_1$. Converting $\mathcal{P}_2$ into the epigraph form, we further obtain

\begin{equation}
\begin{aligned}
& \mathcal{P}_3: && \underset{\mathbf{F_{FM}}}{\text{min}} \; \epsilon \\
& \text{\it s.t.}
& & {\rm C1}: {\norm{|{\bf A}_k {\bf f}_k^{\bf FM}| - {\bf d}(\varphi_k)}}^2_2 \leq \epsilon, \; \forall k \\
&&& {\rm C2}: {\norm{{\bf A}_k {\bf f}_j^{\bf FM}}}^2_2  = 0, \; \forall k, \; j\neq k \\
&&& {\rm C3}: \norm{\mathbf{F_{FM}}}_F^2 \leq K,
\end{aligned}
\end{equation}
where $\epsilon \geq 0$ denotes the maximum matching error between ${\left|{\bf A}_k {\bf f}_k^{\bf FM} \right|}$ and ${{\bf d}(\varphi_k)}$ for all the $K$ RUs. The left-hand side of ${\rm C1}$ can be equivalently expressed as \cite{cao2017constant}

\begin{equation}
 {\norm{\left|{\bf A}_k {\bf f}_k^{\bf FM} \right| - {\bf d}(\varphi_k)}}^2_2 = {\norm{{\bf A}_k {\bf f}_k^{\bf FM} - e^{j {\bm \chi}(\varphi_k)} \odot {\bf d}(\varphi_k)}}^2_2, \; \forall k,
\end{equation}
where ${{\bm \chi}(\varphi_k) = \angle \left({\bf A}_k {\bf f}_k^{\bf FM} \right)}$. We further define the interference matrix ${\bf A}_{{\bf I}_k} \in \mathcal{C}^{\left(K-1\right) N \times M_t}$ given by

\begin{equation}
{\bf A}_{{\bf I}_k}= \left[{\bf A}_1^H, {\bf A}_2^H, \dots, {\bf A}_{k-1}^H, {\bf A}_{k+1}^H, \dots, {\bf A}^H_{K}\right]^H,
\end{equation}
which includes the array response for the other $\left(K-1 \right)$ RUs in the set ${\bm \Phi}_k$. With $ KN \ll M_t$, we obtain ${\rm rank} \left\{{\bf A}_{{\bf I}_k} \right\} = \left( K - 1 \right)N$ and express the singular value decomposition (SVD) of ${\bf A}_{{\bf I}_k}$ as

\begin{equation}
{\bf A}_{{\bf I}_k} = {\bf U}_{{\bf I}_k} {\bf \Sigma}_{{\bf I}_k} {\bf V}^H_{{\bf I}_k},
\end{equation}
where ${\bf V}_{{\bf I}_k} = \left[{\bf v}^k_{{\bf I}_1}, {\bf v}^k_{{\bf I}_2}, \dots, {\bf v}^k_{{\bf I}_{M_t}} \right]$ is the matrix that consists of the right singular vectors. In accordance with the equality constraint ${\rm C2}$ in $\mathcal{P}_3$, we obtain that the precoding vector ${\bf f}_k^{\bf FM}$ is in the null space of ${\bf A}_{{\bf I}_k}$, and can be expressed as a linear combination of the right singular vectors that correspond to zero singular values, given by

\begin{equation}
\begin{split}
\mathbf{f}^{\bf FM}_k &= \sum_{n=1}^{M_t - {\rm rank}\left\{ {\bf A}_{{\bf I}_k} \right\}} \gamma^k_n \cdot {\bf v}^k_{{\bf I}_{   {\rm rank}\left\{ {\bf A}_{{\bf I}_k} \right\} + n  }} \\
&= \sum_{n=1}^{M_t - \left(K - 1 \right)N} \gamma^k_n \cdot {\bf v}^k_{{\bf I}_{   \left(K - 1 \right)N + n  }} \\
&= {{\bf \mathcal{V}}_{{\bf I}_k} }{{\bm \gamma}_k},
\end{split}
\end{equation}
where  ${{\bf \mathcal{V}}_{{\bf I}_k}}  = \left[{\bf v}^k_{{\bf I}_{(K-1)N + 1}}, {\bf v}^k_{{\bf I}_{(K-1)N+2}}, \dots, {\bf v}^k_{{\bf I}_{M_t}} \right]$, each $\gamma^k_n$ represents the weight for the corresponding singular vector and  ${\bm \gamma}_k = \left[\gamma^k_1, \gamma^k_2,  \dots , \gamma^k_{M_t - (K - 1)N}  \right]^T$. Using the expression of $\mathbf{f}^{\bf FM}_k$ in (16) and defining $\mathbf{F_{FM}} = \left[{{\bf \mathcal{V}}_{{\bf I}_1} }{{\bm \gamma}_1}, {{\bf \mathcal{V}}_{{\bf I}_2} }{{\bm \gamma}_2}, \dots, {{\bf \mathcal{V}}_{{\bf I}_K} }{{\bm \gamma}_K} \right]$, $\mathcal{P}_3$ can be formulated into

\begin{equation}
\begin{aligned}
& \mathcal{P}_4: && \underset{{{\bm \gamma}, {\bm \chi}\left(\varphi_{k \in \mathcal{K}} \right)}}{\text{min}} \; \epsilon \\
& \text{\it s.t.}
& & {\rm C1}: {\norm{{\bf A}_k {{\bf \mathcal{V}}_{{\bf I}_k} }{{\bm \gamma}_k} - e^{j {\bm \chi}(\varphi_k)} \odot {\bf d} (\varphi_k)}}^2_2 \leq \epsilon, \; \forall k\\
&&& {\rm C2}: \norm{\mathbf{F_{FM}}}_F^2 \leq K,
\end{aligned}
\end{equation}
where ${\bm \gamma} = \left[{\bm \gamma}_1, {\bm \gamma}_2, \dots, {\bm \gamma}_K \right]$. This problem is not jointly convex with respect to w.r.t. ${\bm \gamma}$ and ${\bm \chi}\left(\varphi_{k \in \mathcal{K}} \right)$ because of the coupling of these variables in the constraint ${\rm C1}$. Nevertheless, for a fixed ${\bm \chi}\left(\varphi_{k \in \mathcal{K}} \right)$, we note that $\mathcal{P}_4$ is convex w.r.t. ${\bm \gamma}$ and is a SOCP, which can be efficiently solved by convex optimization tools \cite{grant2008cvx}. From (13) and (17), for a given ${\bm \gamma}$, we update ${\bm \chi}\left(\varphi_{k \in \mathcal{K}} \right)$ following

\begin{equation}
 { {\bm \chi}(\varphi_k) = \angle \left({\bf A}_k {{\bf \mathcal{V}}_{{\bf I}_k} }{{\bm \gamma}_k} \right)}, \; \forall k.
\end{equation}
Based on the above, $\mathcal{P}_4$ is solved using an alternating optimization process, as shown in Algorithm 1. In Algorithm \ref{DPFM_AG}, the objective function is positive and minimized within each iteration at Step 4 and Step 5. Accordingly, the algorithm converges to a locally optimal solution \cite{cao2017constant, tranter2017fast, rao2009engineering}.

\begin{algorithm}[h]
 \begin{algorithmic}[1]
\STATE \textbf{Input}: $\epsilon_{th}$, $ITR_{max}$, ${ {\bm \chi}(\varphi_k)^{(0)}} \in \left[-\pi, \pi \right], \; \forall k \in \mathcal{K}$
\STATE Initialize $\epsilon_d \gets 1$, $t \gets 0$;
\WHILE {$\epsilon_d \geq \epsilon_{th}$ and $l \leq ITR_{max}$}
\STATE For a given ${{\bm \chi}(\varphi_k)^{(t)}}$,  obtain ${ {\bm \gamma}^{(t+1)}, \epsilon^{(t+1)}}$ by solving $\mathcal{P}_4$;
\STATE Update ${ {\bm \chi}(\varphi_k)^{(t+1)} = \angle \left({\bf A}_k {{{\bf \mathcal{V}}_{{\bf I}_k} }{{\bm \gamma}_k}}^{(t+1)} \right)}, \; \forall k$;
\STATE Update $\epsilon_d = \left|\epsilon^{(t+1)} - \epsilon^{(t)} \right|$; $t \gets t + 1$.
\ENDWHILE
\STATE \textbf{Output}: ${\bf F}^{\bf opt}_{\bf FM} = \left[{{\bf \mathcal{V}}_{{\bf I}_1} }{{\bm \gamma}_1}, {{\bf \mathcal{V}}_{{\bf I}_2} }{{\bm \gamma}_2}, \dots, {{\bf \mathcal{V}}_{{\bf I}_K} }{{\bm \gamma}_K} \right]$
\end{algorithmic}
\caption{Alternating Optimization for the Robust Fully-Digital Precoder in $\mathcal{P}_4$}
\label{DPFM_AG}
\vspace{-0.35em}
\end{algorithm}



\subsection{Robust Fully-Digital Precoder Design based on `Error Statistics' (DP-ES)}

In this section, the `error statistics' metric is incorporated in the DP design to optimize the array gain in the presence of beam misalignment which compared to the `flat-mainlobe' metric introduced in Section III-A can be expressed in a closed form, as detailed below. Based on (4), the array response in the presence of beam alignment error $\delta$ is given by 

\begin{equation}
{{\bm \alpha^e}\left(\hat{\theta} \right)} = \sqrt{\frac{1}{M_t}} \left[1, \alpha^e_1(\hat{\theta}), ...,  \alpha^e_{M_t}(\hat{\theta}) \right]^T,
\end{equation}
where $\alpha^e_m(\hat{\theta}) = \mathbb{E}\left[\alpha_m(\theta + \delta)\right], \; \forall m$, and is computed as

\begin{equation}
\begin{split}
\alpha^e_m(\hat{\theta}) &= \int_{-\beta}^{\beta} e^{j \pi (m-1) \cos(\theta + \delta)} f(\delta) d\delta \\
&= \frac{1}{2 \beta} \int_{-\beta}^{\beta} e^{a_m \cos \delta - b_m \sin \delta} d\delta,
\end{split}
\end{equation}
where $a_m = j \pi (m-1) \cos \theta$, $b_m = j \pi  (m-1) \sin \theta$. Since $\beta$ is small, using Maclaurin series, we have  

\begin{equation}
\begin{split}
a_m \cos(\delta) - b_m \sin(\delta) &= a_m - b_m \delta - \frac{a_m}{2} \delta^2 + \frac{b}{6} \delta^3 + \frac{a}{24} \delta^4 - \frac{b}{120} \delta^5 + \mathcal{O}(\delta^6).
\end{split}
\end{equation}
%
Using the Maclaurian series for the exponential function, we can further express

\begin{equation}
e^{a_m \cos \delta - b_m \sin \delta} = e^{a_m} \sum_{n=0}^5 A_n \delta^n + \mathcal{O}(\delta^6),
\end{equation}
where
\begin{equation}
\begin{split}
        A_0 &= 1, A_1 = -b_m,  A_2 = \frac{1}{2} \left[b_m^2 - a_m \right], A_3 = \frac{1}{6} \left[\left(3a_m - 1\right) b_m -  b_m^3  \right], \\
        A_4 &= \frac{1}{24} \left[\left(3a_m+1 \right) a_m -2 \left(3a_m+2 \right)b_m^2 +  b_m^4 \right],\\
        A_5 &= -\frac{1}{120} \left[15\left(a_m + 1 \right) a_m - 10 \left(a_m+1 \right)b_m^2 + b_m^4 + 1 \right].
\end{split}
\end{equation}
Hence, (20) can now be simplified as

\begin{equation}
\begin{split}
\alpha^e_m(\hat{\theta}) &= \frac{1}{2 \beta} \int_{-\beta}^{\beta} e^{a_m} \sum_n A_n \delta^n  d\delta \\
&= \frac{e^{a_m}}{2 \beta}  \sum_n A_n \frac{\beta^{n+1} - (-1)^{n+1} \beta^{n+1}}{n+1}.
	\end{split}
\end{equation}
By defining $k = 2n + 1, \; n = 0,1,2,...$ and with some algebraic manipulations, the above expression can be recast as

\begin{equation}
\begin{split}
\alpha^e_m(\hat{\theta}) &= e^{a_m} \sum_k A_k \frac{\beta^k}{k+1}.
\end{split}
\end{equation}

We denote the robust fully-digital precoder based on `error statistics' as ${\bf F_{ES}} = \big[{\bf f}_1^{\bf ES}, {\bf f}_2^{\bf ES}, ..., {\bf f}_K^{\bf ES}\big]$, where $\mathbf{f}_k^{\bf ES}$ is the precoder for the $k$-th RU.  Using the principle of zero-forcing \cite{spencer2004zero}, we propose to maximize the array gain for the $k$-th RU subject to zero inter-receiver interference, and the optimization problem on the  fully-digital precoder ${\bf f}_k^{\bf ES}$ in the case of imperfect beam alignment is given by

\begin{equation}
\begin{aligned}
& \mathcal{P}_5: && \underset{\mathbf{f}_k^{\bf ES}}{\text{max}} \;  \left|{{\bm \alpha}^H_{{\bm e}_k} {\bf f}_k^{\bf ES}} \right|^2 \\
& \text{\it s.t.}
& & {\rm C1}: {\bf A}_{{\bf e}_k} {\bf f}_k^{\bf ES} = {\bf 0}, \; \forall k\\
\end{aligned}
\end{equation}
where $\left|{{\bm \alpha}^H_{{\bm e}_k} {\bf f}_k^{\bf ES}}\right|$ is the transmit array gain towards the $k$-th RU, ${ {\bm \alpha}_{{\bm e}_k} = {\bm \alpha^e}\left (\hat{\theta}_k^{(AoD)} \right)}$, $\hat{\theta}_k^{(AoD)}$ is the estimated AoD at the BS for the $k$-th RU and  ${\rm C1}$ cancels the inter-receiver interference. The interference matrix ${\bf A}_{{\bf e}_k} \in \mathcal{C}^{\left(K-1\right) \times M_t}$  is given by

\begin{equation}
{\bf A}_{{\bf e}_k}= \left[{\bm \alpha}_1^{\bf e}, {\bm \alpha}_2^{\bf e}, \dots, {\bm \alpha}_{k-1}^{\bf e}, {\bm \alpha}_{k+1}^{\bf e} \dots, {\bm \alpha}^{\bf e}_{K}\right]^H, 
\end{equation}
which includes the array response for the other $\left(K-1 \right)$ RUs. As $ K \ll M_t$, we have ${\rm rank} \left\{{\bf A}_{{\bf e}_k} \right\} = \left( K - 1 \right) $ and express the singular value decomposition (SVD) of ${\bf A}_{{\bf e}_k}$ as

\begin{equation}
{\bf A}_{{\bf e}_k} = {\bf U}_{{\bf e}_k} {\bf \Sigma}_{{\bf e}_k} {\bf V}^H_{{\bf e}_k},
\end{equation}
where ${\bf V}_{{\bf e}_k} = \left[{\bf v}^k_{{\bf e}_1}, {\bf v}^k_{{\bf e}_2}, \dots, {\bf v}^k_{{\bf e}_{M_t}} \right]$ is the matrix that consists of the right singular vectors. Similar to the previous section, the precoding vector ${\bf f}_k^{\bf ES}$ can be expressed as a linear combination of the right singular vectors of ${\bf A}_{{\bf e}_k}$ that correspond to zero singular values, given by

\begin{equation}
\begin{split}
\mathbf{f}^{\bf ES}_k &= \sum_{n=1}^{M_t - {\rm rank}\left\{ {\bf A}_{{\bf e}_k} \right\}} \beta^k_n \cdot {\bf v}^k_{{\bf e}_{   {\rm rank}\left\{ {\bf A}_{{\bf e}_k} \right\} + n  }} \\
&= \sum_{n=1}^{M_t - K + 1} \beta^k_n \cdot {\bf v}^k_{{\bf e}_{   \left(K - 1 \right) + n  }} \\
&= {{\bf \mathcal{V}}_{{\bf e}_k} }{{\bm \beta}_k},
\end{split}
\end{equation}
where  ${{\bf \mathcal{V}}_{{\bf e}_k}}  = \left[{\bf v}^k_{{\bf e}_{K}}, {\bf v}^k_{{\bf e}_{K+1}}, \dots, {\bf v}^k_{{\bf e}_{M_t}} \right]$, each $\beta^k_n$ represents the weight for the corresponding singular vector, and  ${\bm \beta}_k = \left[\beta^k_1, \beta^k_2,  \dots , \beta^k_{M_t - K + 1}  \right]^T$. Using the expression of $\mathbf{f}^{\bf ES}_k$ in (29), $\mathcal{P}_5$ can be formulated into


\begin{equation}
\begin{aligned}
& \mathcal{P}_6: && \underset{{\bm \beta}_k}{\text{max}} \; \left|{{{\bm \alpha}^H_{{\bm e}_k} {\bf  \mathcal{V}}_{{\bf e}_k} {\bm \beta}_k}}\right|^2, \\
\end{aligned}
\end{equation}
which has the following optimal solution:

\begin{equation}
\begin{split}
 {\bm \beta}_k = {\bf \mathcal{V}}_{{\bf e}_k}^H {\bm \alpha}_{{\bm e}_k}  
\end{split}
\end{equation}
Accordingly, the robust fully-digital precoder for $K$ RUs is given by

\begin{equation}
{\bf F}_{\bf ES} = \left[ {\bf \mathcal{V}}_{{\bf e}_1} {\bf \mathcal{V}}_{{\bf e}_1}^H {\bm \alpha}_{{\bm e}_1}, {\bf \mathcal{V}}_{{\bf e}_2} {\bf \mathcal{V}}_{{\bf e}_2}^H {\bm \alpha}_{{\bm e}_2}, \dots, {\bf \mathcal{V}}_{{\bf e}_K} {\bf \mathcal{V}}_{{\bf e}_K}^H {\bm \alpha}_{{\bm e}_K} \right].
\end{equation}
To satisfy the transmit power constraint at the BS, i.e. $\norm{\mathbf{F_{ES}}}_F^2 \leq K$, we scale ${\bf F}_{\bf ES}$ to obtain the optimal robust fully-digital precoder based on the `error statistic' metric, given by

\begin{equation}
{\bf F}^{\bf opt}_{\bf ES} = \frac{\sqrt{K}}{\norm{{\bf F}_{\bf ES}}_F} {\bf F}_{\bf ES}.
\end{equation}


{\it Discussion}: The major difference between the precoding design based on `flat mainlobe' model and  `error statistics' model is that the former design maximizes the array gain over an angular range around $\hat{\theta}_k^{(AoD)}$ for each RU, while the latter maximizes the array gain only along $\hat{\theta}_k^{(AoD)}, \; \forall k$ by using the expected array response in the presence of the beam  alignment error. Accordingly, while the `flat mainlobe' model aims to maximize the effective array gain, the `error statistics' metric maximizes the average array gain over the misalignment error range for all the RUs and enjoys a closed-form expression for the robust fully-digital precoder.



\subsection{Hybrid Precoding Approximation (HPA)}

Based on the obtained robust fully-digital precoder, in this section we introduce the design of the hybrid precoder $\mathbf{F_{RF}}$ and $\mathbf{F_{BB}}$ based on matrix factorization and alternating optimization. Alternating optimization is widely employed in  optimization problems involving different subsets of variables, which also finds its  applications  in  matrix completion \cite{jain2013low}, \cite{netrapalli2013phase}, image reconstruction \cite{wang2008new},  blind  deconvolution  \cite{chan2000convergence}  and  non-negative  matrix factorization \cite{kim2008nonnegative}.  Similar to the work in \cite{yu2016alternating}, we design the hybrid precoding by formulating the following optimization:

\begin{equation}
\begin{aligned}
& \mathcal{P}_7: && \underset{\mathbf{\mathbf{F_{RF},F_{BB}}}}{\text{min}} \; \norm{\mathbf{F_{opt} - F_{RF}F_{BB}}}^2_F \\
& \text{\it s.t}
& & {\rm C1}: \norm{\mathbf{F_{RF} F_{BB}}}_F^2 \leq K \\
&&& {\rm C2}: \left| \left[\mathbf{F_{RF}} \right]_{m,n} \right| = \sqrt{\frac{1}{M_t}}, \; \forall m,n ,  
\end{aligned}
\end{equation}
which is a  matrix  factorization  problem and is solved by alternately optimizing $\mathbf{F_{RF}}$ and $\mathbf{F_{BB}}$. Algorithm \ref{HADP_AG} describes the framework to obtain a feasible hybrid precoder by approximating the fully-digital precoder based on the principle of alternating optimization, as shown below. To be more specific, the digital precoder $\mathbf{F_{BB}}$ is designed based on a fixed analog precoder $\mathbf{F_{RF}}$ by solving the following least-square sub-problem:

\begin{equation}
\begin{aligned}
& \mathcal{P}_8: && \underset{\mathbf{\mathbf{F_{BB}}}}{\text{min}} \; \norm{\mathbf{F_{opt} - F_{RF}F_{BB}}}^2_F, \\
& \text{\it s.t}
& & {\rm C1}: \norm{{\bf F_{RF}} {\bf F_{BB}}}_F^2 \leq K \\
\end{aligned}
\end{equation}
which leads to

\begin{equation}
\mathbf{F_{BB}} = {\bf F}_{\bf RF}^{\dagger} {\bf F}_{\bf opt},
\end{equation}
where ${\rm C1}$ in $\mathcal{P}_8$ is satisfied by normalizing $\mathbf{F_{BB}}$ by the factor $\frac{\sqrt{K}}{\norm{{\bf F_{RF} F_{BB}}}_F}$ \cite[Lemma 1]{yu2016alternating}.

\begin{algorithm}[h]
 \begin{algorithmic}[1]
\STATE \textbf{Input}: $\mathbf{F_{opt}}$, $\epsilon_{th}$, $ITR_{max}$
\STATE Initialize $\epsilon^{(0)} \gets 0$, $\epsilon_d \gets 0$, $t \gets 0$;
\STATE Initialize ${\bf F}_{\bf RF}^{(0)}$ using random phase;
\WHILE {$\epsilon_d \geq \epsilon_{th}$ and $t \leq ITR_{max}$}
\STATE For a given ${{\bf F}^{(t)}_{\bf RF}}$, calculate ${{\bf F}^{(t+1)}_{\bf BB}} = \left( {{\bf F}^{(t)}_{\bf RF}} \right)^{\dagger} {\bf F_{opt}}$;
\STATE Update ${{\bf F}^{(t+1)}_{\bf RF}}$;
\STATE $\epsilon^{(t+1)} = \norm{{{\bf F_{opt}} - {{\bf F}^{(t+1)}_{\bf RF}} {{\bf F}^{(t+1)}_{\bf BB}} }}_F^2$;
\STATE $\epsilon_d = \left|\epsilon^{(t+1)} - \epsilon^{(t)} \right|$; $t \gets t + 1$.
\ENDWHILE
\STATE ${{\bf F_{BB}} = \frac{\sqrt{K}}{\norm{{\bf F_{RF} F_{BB}}}_F} {\bf F_{BB}}}$.
\STATE \textbf{Output}: $\mathbf{F}^{\bf opt}_{\bf RF}, \mathbf{F}^{\bf opt}_{\bf BB}$
\end{algorithmic}
\caption{Hybrid Precoding by Approximating the Fully-Digital Precoder}
\label{HADP_AG}
\vspace{-0.35em}
\end{algorithm}

The analog precoder $\mathbf{F_{RF}}$ for a given $\mathbf{F_{BB}}$ is designed by solving the following sub-problem:

\begin{equation}
\begin{aligned}
& \mathcal{P}_9: && \underset{\mathbf{\mathbf{F_{RF}}}}{\text{min}} \; \norm{\mathbf{F_{opt} - F_{RF}F_{BB}}}^2_F \\
& \text{\it s.t.}
& & {\rm C1}: \left| \left[\mathbf{F_{RF}} \right]_{m,n} \right| = \sqrt{\frac{1}{M_t}}, \; \forall m,n, 
\end{aligned}
\end{equation}
By defining $\mathbf{f} = vec\big({\bf F_{opt}\big)}$, ${{\bf A}_{\eta} = {\bf F}_{\bf BB}^T \otimes {\bf I}_{M_t}}$ and  $\mathbf{x} = vec\big({\bf F_{RF}} \big)$, $\mathcal{P}_9$ is reformulated as the following constant modulus least-square (CMLS) problem:

\begin{equation}
\begin{aligned}
& \mathcal{P}_{10}: && \underset{\mathbf{\mathbf{x}}}{\text{min}} \; \norm{\mathbf{f - A_{\eta} x}}^2_2 \\
& \text{\it s.t.}
& & {\rm C1}: \big|[{\bf x}]_{n}\big| = \sqrt{\frac{1}{M_t}}, \; n = 1,2,...M_t N_{RF},  
\end{aligned}
\end{equation}
In the following, we first briefly review an existing algorithm for $\mathcal{P}_{10}$ based on manifold optimization, followed by the description for our proposed algorithms based on GP and LSP.

\subsubsection{Analog Precoding Design - Manifold Optimization (MO) \cite{yu2016alternating}}

For the MO as a benchmark in this paper, the constraint ${\rm C1}$ in $\mathcal{P}_{10}$ is defined as a Riemannian manifold \cite{absil2009optimization}. Endowing the complex plane $\mathcal{C}$ with the Euclidean metric $<x_1,x_2> = \mathcal{R}\{x_1^* x_2\}$, the manifold $\mathcal{M}_c =\{\mathbf{x} \in \mathcal{C}^n: |x_1| = 1, |x_2| = 1, ..., |x_n| = 1 \}$ where $n = M_t N_{RF}$ is introduced, which is the search space of $\mathcal{P}_{10}$ over Riemannian submanifold of  $\mathcal{C}^n$.

Defining the tangent space for a point $\mathbf{x} \in \mathcal{M}_c$ as $\mathcal{T}_x \mathcal{M}_c = \{\mathbf{t} \in \mathcal{C}^{n}: \mathcal{R} \{\mathbf{t \odot x }\} = 0 \}$, the Riemannian gradient $grad {\bm \xi(x)}$ is obtained by the orthogonal projection of the Euclidean gradient ${\nabla \bm {\xi(x)} \in \mathcal{C}}^n$ onto $\mathcal{T}_x \mathcal{M}_c$, given as

\begin{equation}
  \begin{split}
      grad \mathbf{{\bm \xi}(x)} &= \mathbf{Proj_x \nabla {\bm \xi(x)}},\\
      &= \mathbf{\nabla {\bm \xi}(x)} -  \mathcal{R}\{\mathbf{\nabla {\bm \xi}(x)} \odot \mathbf{x}^* \} \odot \mathbf{x},
\end{split}
\end{equation}
where for the unconstrained objective of $\mathcal{P}_{10}$, $\mathbf{\nabla {\bm \xi}(x)}$  is given by 

\begin{equation}
 \mathbf{\nabla {\bm \xi} \left(x \right)} = -2 {\bf A}^H_{\eta} \left({\bf f} - {\bf A}_{\eta} {\bf x} \right).
\end{equation}

To evaluate the objective function on the manifold, the retraction operation $R_\mathbf{x}$ of a tangent vector $\mathbf{d}$ at the point $\mathbf{x \in \mathcal{M}}_c$ is defined as \cite{boumal2014manopt}

\begin{equation}
R_{\mathbf{x}}: \mathcal{T}_{\mathbf{x}}\mathcal{M}_c \rightarrow \mathcal{M}_c: \mathbf{\alpha d} \mapsto R_{\mathbf{x}} = vec \bigg[ \frac{(\mathbf{x + \alpha d})_i}{\big|(\mathbf{x + \alpha d})_i \big|}\bigg].
\end{equation}

The counterpart of the classical conjugate gradient method (CGM) for the defined manifold is used to search for the optimal analog precoder in \cite{yu2016alternating}, where it is observed  that updating the Riemannian gradient and the descent direction requires the operation between two vectors in different tangent spaces $\mathcal{T}_{\mathbf{x}_i} \mathcal{M}_c$ and $\mathcal{T}_{\mathbf{x}_{i+1}} \mathcal{M}_c$. Subsequently, the tangent vector $\mathbf{d}$ needs to be mapped from $\mathbf{x}_i$ to $\mathbf{x}_{i+1}$, which is given by

\begin{equation}
Tp_{\mathbf{x}_i \rightarrow {\bf x}_{i+1}}: \mathcal{T}_{\mathbf{x}_i}\mathcal{M}_c \rightarrow  \mathcal{T}_{\mathbf{x}_{i+1}}\mathcal{M}_c: \mathbf{d} \mapsto \mathbf{d} - \mathcal{R} \{{{\bf d} \odot {\bf x}^*_{i+1} \} \odot \mathbf{x}_{i+1}}.
\end{equation}

While the MO algorithm can provide a near-optimal performance, the  update of  the  analog precoder  involves a  line-search  algorithm, which within each iteration involves a) an orthogonal projection of Euclidean gradient $grad {\bm \xi(\bf x)}$  onto the tangent space defined in (39), b) a retraction of the tangent vector on the manifold  $R_{\mathbf{x}}$ defined in (41), and c) the construction of  a transport from one tangent space to another, as defined in (42).

\subsubsection{Proposed Analog Precoding Design - Gradient Projection (GP)}

In this section,  we propose an iterative analog precoding design based on the GP method, which only requires the projection of the solution sequence onto the element-wise constant modulus constraint set. The GP method is in nature a revamped version of the Conjugate  Gradient Method (CGM), which searches for the optimal solution by projecting each subsequent point
\begin{equation}
{\mathcal{X}^{(t+1)} = {\bf x}^{(t)}} + \alpha_{GP} \mathbf{d}^{(t)}
\end{equation}
onto the feasible region ${\rm C1}$ defined in $\mathcal{P}_{10}$, while moving along the decent direction ${\bf d}^{(t)} = -{\nabla {\bf f} \left({\bf x}^{(t)} \right)}$, with the step size $\alpha_{GP}$ given by

\begin{equation}
\begin{split}
\alpha_{GP} &= \argmin_{\alpha \geq 0} \norm{{{\bf f} - {\bf A}_{\eta} {\bf x}}}^2_2 \bigg|_{{{\bf x}^{(t)}} + \alpha {\bf d}^{(t)}} \\
&= {\left[ \left({{\bf d}}^{(t)} \right)^H {\bf A}_{\eta}^H {\bf A}_{\eta} {\bf d}^{(t)} \right]^{-1} \bigg(\mathcal{R} \left\{ {\bf f}^H {\bf A}_{\eta} {\bf d}^{(t)} \right\} - \mathcal{R} \left \{ \left({  {\bf d}^{(t)}} \right)^H {\bf A}_{\eta}^H {\bf A}_{\eta} {\bf x}^{(t)} \right\}} \bigg).
\end{split}
\end{equation}
The projection onto ${\rm C1}$ can be viewed as a phase-extraction operation and has a closed-form solution, given by

\begin{equation}
 {{\bf x}^{(t+1)} = \sqrt{\frac{1}{M_t}} e^{j \angle {\bf \mathcal{X}}^{(t+1)}}}.
\end{equation}

Algorithm \ref{GP_AG} summarizes the GP method, which employs the Polak-Ribiere parameter $\beta_{GP}$  \cite{rao2009engineering} to update the decent direction in each iteration, as shown in Step 9 of Algorithm \ref{GP_AG}. \cite{tranter2017fast} has demonstrated that the projection onto the element-wise constant modulus constraint set does not increase the objective, as the solution sequence $\mathbf{x}^{(t)}$ converges to a KKT point. 

\begin{algorithm}[h]
 \begin{algorithmic}[1]
\STATE \textbf{Input}: $\epsilon_{th}$, $ITR_{max}$
\STATE Initialize $\mathbf{x}^{(0)}$ with random phase, $\mathbf{d}^{(0)} \gets -{\nabla {\bm \xi} \left({\bf x}^{(0)} \right)}$,   $\epsilon^{(0)} \gets 0$, $\epsilon_d \gets 0$, $t \gets 0$;
\WHILE {$\epsilon_d \geq \epsilon_{th}$ and $t \leq ITR_{max}$}
\STATE $\alpha_{GP} = {\left[ \left({{\bf d}}^{(t)} \right)^H {\bf A}_{\eta}^H {\bf A}_{\eta} {\bf d}^{(t)} \right]^{-1} \bigg(\mathcal{R} \left\{ {\bf f}^H {\bf A}_{\eta} {\bf d}^{(t)} \right\} - \mathcal{R} \left \{ \left({  {\bf d}^{(t)}} \right)^H {\bf A}_{\eta}^H {\bf A}_{\eta} {\bf x}^{(t)} \right\}} \bigg)$;
\STATE ${\mathcal{X}^{(t+1)} = {\bf x}^{(t)}} + \alpha_{GP} \mathbf{d}^{(t)}$;
\STATE ${{\bf x}^{(t+1)} = \sqrt{\frac{1}{M_t}} e^{j \angle {\bf \mathcal{X}}^{(t+1)}}}$;
\STATE ${\nabla {\bf {\bm \xi}} \left({\bf x}^{(t+1)} \right)} = -2 {\bf A}^H_{\eta} \left({\bf f} - {\bf A}_{\eta} {\bf x}^{(t+1)} \right)$; 
\STATE $\beta_{GP} = \frac{ \left({\nabla {\bm \xi} \left({\bf x}^{(t+1)} \right)} - {\nabla {\bm \xi} \left({\bf x}^{(t)} \right)} \right)^H {\nabla {\bm \xi} \left({\bf x}^{(t+1)} \right)}}{ \left({\nabla {\bm \xi} \left({\bf x}^{(t)} \right)} \right)^H {\nabla {\bm \xi} \left({\bf x}^{(t)} \right)} }$;
\STATE ${\bf d}^{(t+1)} = - {\nabla {\bf f} \left({\bf x}^{(t+1)} \right)} + \beta_{GP} \mathbf{d}^{(t)}$;
\STATE $\epsilon^{(t+1)} = \norm{{{\bf f} - {\bf A}_{\eta} {\bf x}^{(t)}}}_F^2$; 
\STATE $\epsilon_d = \left| \epsilon^{(t+1)} - \epsilon^{(t)} \right| $; $t \gets t + 1$.
\ENDWHILE
\STATE \textbf{Output}: ${{\bf x}^{(t+1)}}$
\end{algorithmic}
\caption{Analog Precoding Design - Gradient Projection (GP)}
\label{GP_AG}
\vspace{-0.35em}
\end{algorithm}

\subsubsection{Proposed Low-Complexity Analog Precoding Design - Least Square Projection (LSP)}

Although the proposed GP method requires a lower computational cost compared to the MO algorithm, as will be analyzed and numerically shown in the following, it still involves a line-search process. Accordingly, a large variable size may significantly slow down the convergence speed and prevent its practical implementation. Therefore in this section, we propose an alternative to the GP method for the analog precoding design that does not require an iterative process. To be more specific, we observe that the unconstrained least-square problem $\mathcal{P}_{10}$ has a well-defined solution, given by \cite{rao2009engineering}

\begin{equation}
\mathbf{\tilde{x} = A_{\eta}^H \big(A_{\eta} A_{\eta}^H \big)^{-1} f}.
\end{equation}
Based on the properties of Kronecker product, we further obtain   

\begin{equation}
\begin{split}
  {{\bf A}_\eta {\bf A}_\eta^H} &=  {\big({\bf F}^T_{\bf BB} \otimes {\bf I}_{M_t}\big) \big({\bf F}^T_{\bf BB} \otimes {\bf I}_{M_t} \big)^H} \\
                &=  {\big({\bf F}^T_{\bf BB} \otimes {\bf I}_{M_t}\big) \big({\bf F}^*_{\bf BB} \otimes {\bf I}_{M_t}\big)} \\
                &=  {{\bf F}^T_{\bf BB} {\bf F}^*_{\bf BB} \otimes {\bf I}_{M_t} {\bf I}_{M_t}} =  \alpha^2 {{\bf I}_{K} \otimes {\bf I}_{M_t} {\bf I}_{M_t}} \\
               &=   \alpha^2 {{\bf I}_{K M_t}}, \\
\end{split}
\end{equation}
where  ${\bf F}_{\bf BB}^T {\bf F}_{\bf BB}^* = \alpha^2 {\bf I}_K$ is imposed due to the orthogonality of the full-digital precoder ${\bf F}_{\bf opt}$ to mitigate the inter-receiver interference \cite{yu2016alternating} which further leads to $\mathbf{\tilde{x}} = \frac{1}{\alpha^2} {\bf A}_{\eta}^H {\bf f}$. To satisfy the constraint in $\mathcal{P}_{10}$, we directly project the solution to the unconstrained $\mathcal{P}_{10}$ onto the constant modulus space and hence ignoring the constant $\frac{1}{\alpha^2}$, we have:

\begin{equation}
\mathbf{x} = \sqrt{\frac{1}{M_t}} e^{j \angle \left({\bf A}_{\eta}^H {\bf f} \right)},
\end{equation}
and the analog precoding matrix can be accordingly obtained, which avoids the complex matrix inversion and the iterative process.

\subsection{Computational Complexity Analysis for Analog Precoding Design}

In this section, we study the computational costs of the above several analog precoding designs in terms of the floating-point operations required. We evaluate the computational complexity in terms of the required number of multiplications/divisions and additions/subtractions for each analog precoding design.

\begin{enumerate}
\item {\bf MO Method:} The complexity of the gradient-search process is dominated by the line-search process and the computation of the gradient. In \cite{yu2016alternating}, the MO method deploys the Armijo backtracking line search, where each iteration involves the computation of the cost function, i.e. objective of $\mathcal{P}_{10}$, whose complexity is given by

\begin{equation}
   C_{ALS} =  \eta_{LS} \left( 2 M_t^2 N_{RF} K + 4 M_t K + M_t^2 N^2_{RF} + 2 M_t N_{RF} \right),
\end{equation}
where $\eta_{LS}$ is the number of iterations required for the line search. The computation of the subsequent point and its retraction from the tangent space onto the manifold, as defined in (41), requires the complexity of 

\begin{equation}
   C_{R_{\bf x}} = M_t N_{RF},
\end{equation}
and the computation of the Riemannian gradient given by (39) consumes

\begin{equation}
   C_{P_{\bf \nabla f}} = 3 M_t N_{RF}.
\end{equation}
The complexity for transporting the Riemannian gradient and the descent direction between different tangent spaces in (42) is given by 

\begin{equation}
   C_{T{\bf x}} = 6 M_t N_{RF},
\end{equation}
while the computation of the Polak-Ribiere parameter and the update of the decent direction require the complexity of

\begin{subequations}
\begin{align}
        C_{PR} &= 3 M_t N_{RF},\\
        C_{U_d} &= M_t N_{RF},
\end{align}
\end{subequations}
respectively. Hence, assuming a maximum number of iterations $\eta_{max}^{MO}$, the overall of complexity for the MO can be expressed as

\begin{equation}
\begin{split}
    C_{MO} &=  \eta_{max}^{MO} \left\{C_{ALS} + C_{R_{\bf x}} + C_{P_{\bf \nabla f}} + C_{T{\bf x}} +  C_{PR} + C_{U_d} \right\} \\
           &=   2\eta_{max}^{MO} M_t \left[ 2\eta_{LS} \left( M_t N_{RF} + 2 \eta_{LS} \right) K + \left(\eta_{LS} + 1\right) M_t N_{RF}^2 + \left(2 \eta_{LS} + 14 \right) N_{RF} \right].
\end{split}
\end{equation}

\item{\bf GP Method:} In GP, we use the exact line search (ELS) given by (44) whose computation is dominated by evaluation of ${\bf A_{\eta} {\bf d}^{(t)}}$, where ${\bf A}_{\eta} \in \mathcal{C}^{M_t K \times M_t N_{RF}}$ and ${\bf d}^{(t)} \in \mathcal{C}^{M_t N_{RF} \times 1}$, which has a overall complexity of 

\begin{equation}
   C_{ELS} = 5 M_t^2 N_{RF} K + M_t K + M_t N_{RF}.
\end{equation}
The computation of the subsequent point and the projection onto of the constraint ${\rm C1}$ of $\mathcal{P}_{10}$, as defined in (45), require the complexity of 

\begin{equation}
   C_{P_{GP}} =  2 M_t N_{RF}.
\end{equation}

The complexity for updating the gradient of the unconstrained objective of  $\mathcal{P}_{10}$ is given by 
 
 \begin{equation}
   C_{G_{\nabla f}} = 2 M_t^2 N_{RF} K + M_t K,
\end{equation}
and the complexity to calculate Polak-Ribiere parameter and update the decent direction is the same as that shown in (53). Hence, assuming a maximum number of iterations $\eta_{max}^{GP}$, the overall complexity of the GP method is given by

\begin{equation}
\begin{split}
    C_{GP} &=  \eta_{max}^{GP}\left\{ C_{ELS} + C_{P_{GP}} + C_{G_{\bf \nabla f}} + C_{PR} + C_{U_d} \right\} \\
           &=  M_t \eta_{max}^{GP}  \left[7 N_{RF} \left(M_t K + 1 \right) + 2 K \right].
           \end{split}
\end{equation}

\item{\bf LSP Method:} The complexity of the LSP method is dominated by the computation of ${\bf A}_{\eta}^H {\bf f}$, where ${\bf A}_{\eta} \in \mathcal{C}^{M_t K \times M_t N_{RF}}$ and ${\bf f} \in \mathcal{C}^{M_t K \times 1}$, and its projection onto the element wise constant modulus set defined in (48). Hence, the overall complexity of the LSP is given by

\begin{equation}
  C_{LSP} = M_t N_{RF} \left(M_t K + 1 \right).
\end{equation}

\end{enumerate}

To numerically compare the complexity of the schemes, in Table \ref{TCC} we show the number of computations required w.r.t.  the number of RUs $K$, the number of BS antennas $M_t$ and the number of RF chains $N_{RF}$. Based on the simulations, we have taken $\eta_{LS} = 2$ and $\eta_{max}^{MO} = \eta_{max}^{GP} = 100$. As can be seen, the computational cost of the LSP method is significantly lower than the proposed GP method, while GP method requires around $30 \%$ to $40 \%$ of the total computation required by MO.


\begin{table}
\begin{center}
\setlength\arrayrulewidth{1.50pt}
\begin{tabular}{ |p{0.35cm}|p{1.40cm}|p{1.4cm}|p{1.4cm}||p{1.4cm}|p{1.4cm}|p{1.4cm}||p{1.45cm}|p{1.4cm}|p{1.4cm}| }
 \hline
 \multirow{2}{*}{$M_t$} & \multicolumn{3}{|c||}{Schemes, $N_{RF} = 6$, $K = 4$} & \multicolumn{3}{|c||}{Schemes, $N_{RF} = 12$, $K = 4$} & \multicolumn{3}{|c|}{Schemes, $N_{RF} = 12$, $K = 8$} \\
\cline{2-10} 
 {} & MO & GP & LSP & MO & GP & LSP & MO & GP & LSP\\ 
 \hline
 128   & $6.72 \times 10^{8}$  & $2.76 \times 10^{8}$ &  $3.94 \times 10^{5}$ & $2.05 \times 10^{9}$  & $5.52 \times 10^{8}$ &  $7.88 \times 10^{5}$ & $2.68 \times 10^{9}$  & $1.10\times 10^{9}$ &  $1.57 \times 10^{6}$\\
  \hline
  160 &   $1.05 \times 10^{9}$  & $4.31 \times 10^{8}$   &$6.15 \times 10^{5}$ &  $3.20 \times 10^{9}$  & $7.39 \times 10^{8}$   &$1.23 \times 10^{6}$& $4.19 \times 10^{9}$ & $1.72 \times 10^{9}$ & $2.46 \times 10^{6}$\\
  \hline
192 &$1.15 \times 10^{9}$ & $6.20 \times 10^{8}$ &  $8.86 \times 10^{5}$ &    $4.61 \times 10^{9}$  & $1.24 \times 10^{9}$   &$1.77 \times 10^{6}$ & $6.03 \times 10^{9}$ & $2.48 \times 10^{9}$ & $3.54 \times 10^{6}$\\
 \hline
 256 &$2.68 \times 10^{9}$ & $1.10 \times 10^{9}$&  $1.57 \times 10^{6}$  &   $8.19 \times 10^{9}$  & $2.20 \times 10^{9}$   &$3.15 \times 10^{6}$ & $1.07 \times 10^{10}$ & $4.40 \times 10^{9}$ & $6.29 \times 10^{6}$\\
\hline
\end{tabular}
\end{center}
\caption{\small Numerical comparison of the computational costs for different analog precoding schemes, $\eta_{LS} = 2$ and $\eta_{max}^{MO} = \eta_{max}^{GP} = 100$.}
\label{TCC}
\end{table}






\section{Inter-receiver Interference Cancellation}

To mitigate any potential residual 
interference due to the approximation involved in the hybrid precoding design, in this section, we introduce a second-stage digital precoder $\mathbf{F_{BD}}$, which when cascaded with $\mathbf{F_{BB}}$, removes the inter-receiver interference based on zero-forcing.  To be specific, with the obtained hybrid precoder, the effective channel vector for the $k$-th RU is given by

\begin{equation}
{{\bf h}_{{\bf ef}_k} = {\bf w}_k^H {\bf H}_k {\bf F_{RF}} {\bf F_{BB}}}.  
\end{equation}

To fully cancel the inter-receiver interference, the second-stage digital precoder $\mathbf{F_{BD}} = [{\bf f}_1^{\bf BD}, {\bf f}_2^{\bf BD}, ...,{\bf f}_K^{\bf BD}]$ is defined to satisfy the following condition:

\begin{equation}
{{\bf h}_{{\bf ef}_j} {\bf f}_k^{\bf BD} = 0}, \forall k \neq j.
\end{equation}
Following the fully-digital block diagonal precoding \cite{spencer2004zero},  we define the effective channel matrix ${{\bf H}_{{\bf ef}_k}} \in \mathcal{C}^{\left(K-1\right) \times K}$, which consists of all the effective channel vectors except ${{\bf h}_{{\bf ef}_k}}$, i.e. ${{\bf H}_{{\bf ef}_k} = [{\bf h}^T_{{\bf ef}_1},.., {\bf h}^T_{{\bf ef}_{k-1}}, {\bf h}^T_{{\bf ef}_{k+1}}, ...,{\bf h}^T_{{\bf ef}_K}]^T}$, which leads to

\begin{equation}
   {{\bf H}_{{\bf ef}_k} {\bf f}_k^{\bf BD} = 0}, \forall k. 
 \end{equation}
 
 The singular value decomposition (SVD) of ${\bf H}_{{\bf ef}_k}$ is given by 

\begin{equation}
{\bf H}_{{\bf ef}_k} = {\bf U}_{{\bf ef}_k} {\bf \Sigma}_{{\bf ef}_k}  {\bf V}_{{\bf ef}_k}^H,
\end{equation}
where ${\bf V}_{{\bf ef}_k} = \left[{\bf v}^k_{{\bf ef}_1}, {\bf v}^k_{{\bf ef}_2}, \dots, {\bf v}^k_{{\bf ef}_{K}} \right]$ is the matrix that consists of the right singular vectors. According to the equality condition (62), the precoding vector for the $k$-th RU  ${{\bf f}_k^{\bf BD}}$ is in the null space of ${\bf H}_{{\bf ef}_k}$, and consequently the second-stage digital precoder is given by

\begin{equation}
{\bf F_{BD}} = \left[{\bf v}^1_{{\bf ef}_{K}}, {\bf v}^2_{{\bf ef}_{K}}, \dots, {\bf v}^k_{{\bf ef}_{K}}, \dots, {\bf v}^K_{{\bf ef}_{K}} \right], 
\end{equation}
where $ {\bf v}^k_{{\bf ef}_{K}}$ is the $K$-th right column of $\mathbf{V}_{{\bf ef}_k}$. Finally, the $M_t \times K$ composite hybrid precoding matrix ${\bf F}_{\bf HP}  = [{\bf f}_1^{\bf HP}, {\bf f}_2^{\bf HP}, \dots, {\bf f}_K^{\bf HP}]$ is given by

\begin{equation}
\mathbf{F_{HP}} = \frac{\sqrt{K}}{\norm{{\bf F_{RF} F_{BB} F_{BD}}}_F} {\bf F_{RF} F_{BB} F_{BD}}.
\end{equation}
%



\section{Numerical Results}

In  this  section,  we  evaluate  the  performance of our proposed schemes via Monte-Carlo simulations. Unless stated otherwise, we consider a multi-receiver mmWave communication system where the BS  is equipped with $M_t = 128$  ULA antennas and $N_{RF}= 8$ RF chains, and there is a total number of $ K= 4$ RUs, with each equipped with $M_r = 32$ ULA antennas. We assume $L_k = 3, \; \forall k$ for the mmWave channel, and  the beam alignment error with a standard deviation of $\Delta = 1.154$, i.e.,   $-2^o \leq \delta \leq 2^o$, is used. For the `flat-mainlobe' schemes, $\mathbf{\Phi}_k$ contains $N= 8$ samples uniformly distributed in the range of $\hat{\theta}_k^{(AoD)} + \left|\beta\right|,  \; \forall k$. The SNR is defined as $ {\rm SNR} = 10 \; \log_{10} \frac{1}{\sigma^2}$, where the total transmit power is taken as $P = 1$ with uniform power allocation for the $K$ RUs. The  antenna spacing is $d = \frac{\lambda}{2}$  and  all  simulation  results  are  averaged  over  ${\rm 10^3}$ channel realizations. As a benchmark, the conventional fully-digital precoder (CDP) ${\bf f}_k^{\bf D}$ for the $k$-th RU is obtained by solving the following problem:

\begin{equation}
\begin{aligned}
& \mathcal{P}_{11}: && \underset{\mathbf{f}_k^{\bf D}}{\text{max}} \; \left|{\bm \alpha}^H_{k} {\bf f}_k^{\bf D} \right|^2 \\
& \text{\it s.t.}
& & {\rm C1}: {\bf A}_{\bf I} {\bf f}_k^{\bf D} = 0, \\
\end{aligned}
\end{equation}
where ${\bf A_I} = \big[{\bm \alpha}_1, {\bm \alpha}_2, \dots, {\bm \alpha}_{k-1}, {\bm \alpha}_{k+1} \dots, {\bm \alpha}_{K}\big]^H$ with ${ {\bm \alpha}_l = {\bm \alpha}\left (\hat{\theta}_l^{(AoD)} \right)}$, ${\bf F}_{\bf D} = \left[{\bf f}_1^{\bf D}, {\bf f}_2^{\bf D}, \dots, {\bf f}_K^{\bf D} \right]$ and is solved according to the procedure adopted for DP-ES to satisfy the BS power constraint $\norm{\mathbf{F_D}}_F^2 \leq K$. For clarity we define the following abbreviations that are used throughout this section:

\begin{enumerate}

\item FM-MO/GP/LSP: The hybrid precoding design based on `flat-mainlobe' with the analog design based on MO, GP, or LSP;

\item ES-MO/GP/LSP: The hybrid precoding design based on `error statistics' with the analog design based on MO, GP, or LSP;




\item OMP: The hybrid precoding design proposed in \cite{el2014spatially} based on Orthogonal Matching Pursuit Algorithm to approximate CDP;

\item MO-AltMin-CDP: The hybrid precoding design based on MO approximating CDP; 

\end{enumerate}

First, the convergence of the algorithms for the precoder design is validated in Fig.2. In the Fig.2(a), we show the convergence of the GP algorithm for the analog precoder design. It can be seen that for a fixed $\mathbf{F_{BB}}$, GP algorithm converges within $25$ to $30$ iterations for both FM-GP and ES-GP. The convergence of the alternating optimization described in Algorithm 3 is presented in Fig.2(b). Though for all the schemes, the algorithm converges within $10$ iterations, FM-LSP and ES-LSP have a substandard convergence compared to FM-GP and ES-GP. This can be explained from the fact that at each step of Algorithm 3 and with a  fixed  $\mathbf{F_{BB}}$, a near-optimal analog precoder is obtained with the GP algorithm, while the LSP gives only a least-square approximation of the analog precoder. It results in an inferior performance of the LSP based schemes compared to the GP based schemes which is reflected in the subsequent performance results.

\begin{figure}[!htb]
    \centering
   \subfigure[]
    {
        \includegraphics[scale=0.31]{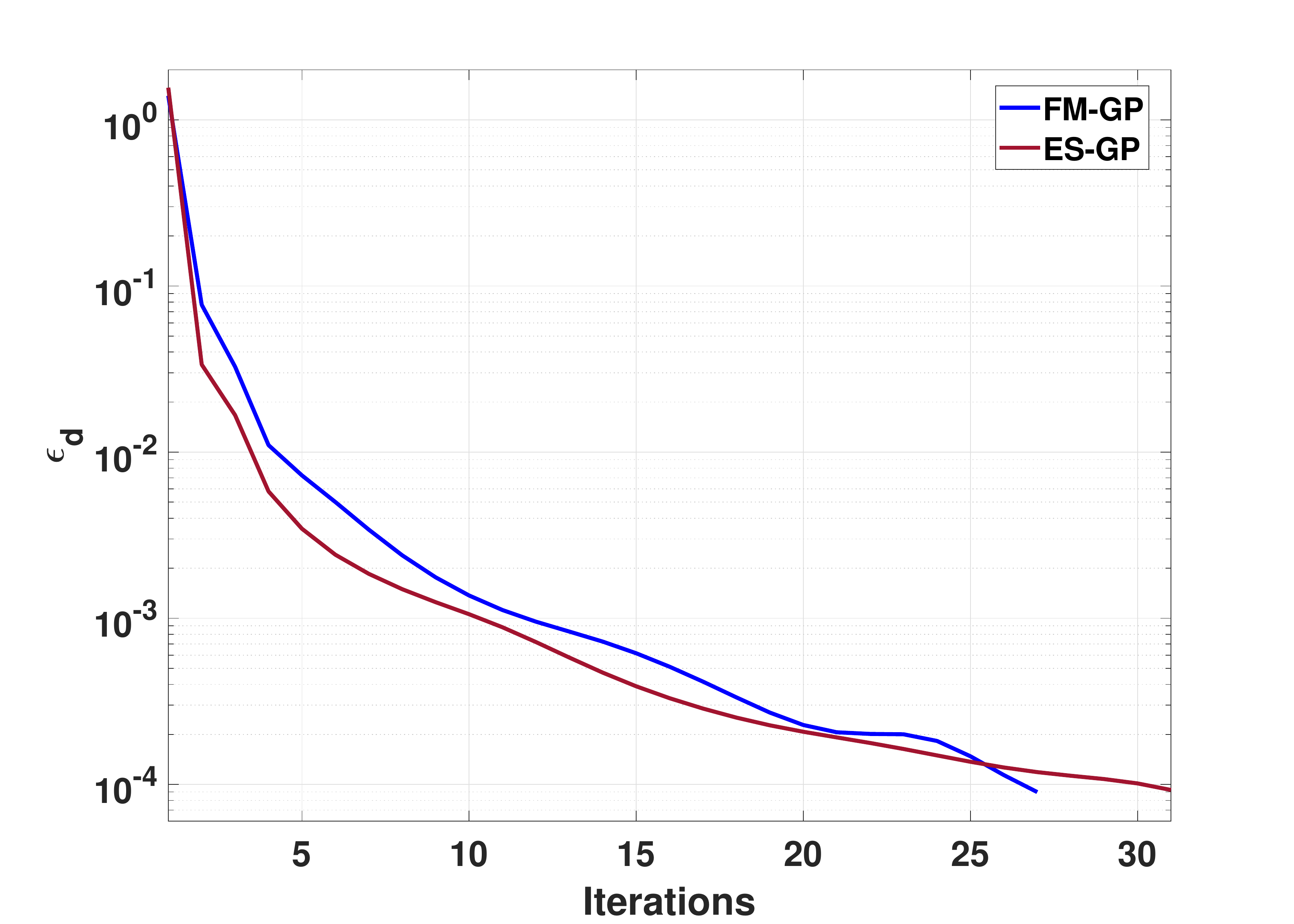}
    } \hskip -4.5ex
    \subfigure[]
    {
         \includegraphics[scale=0.31]{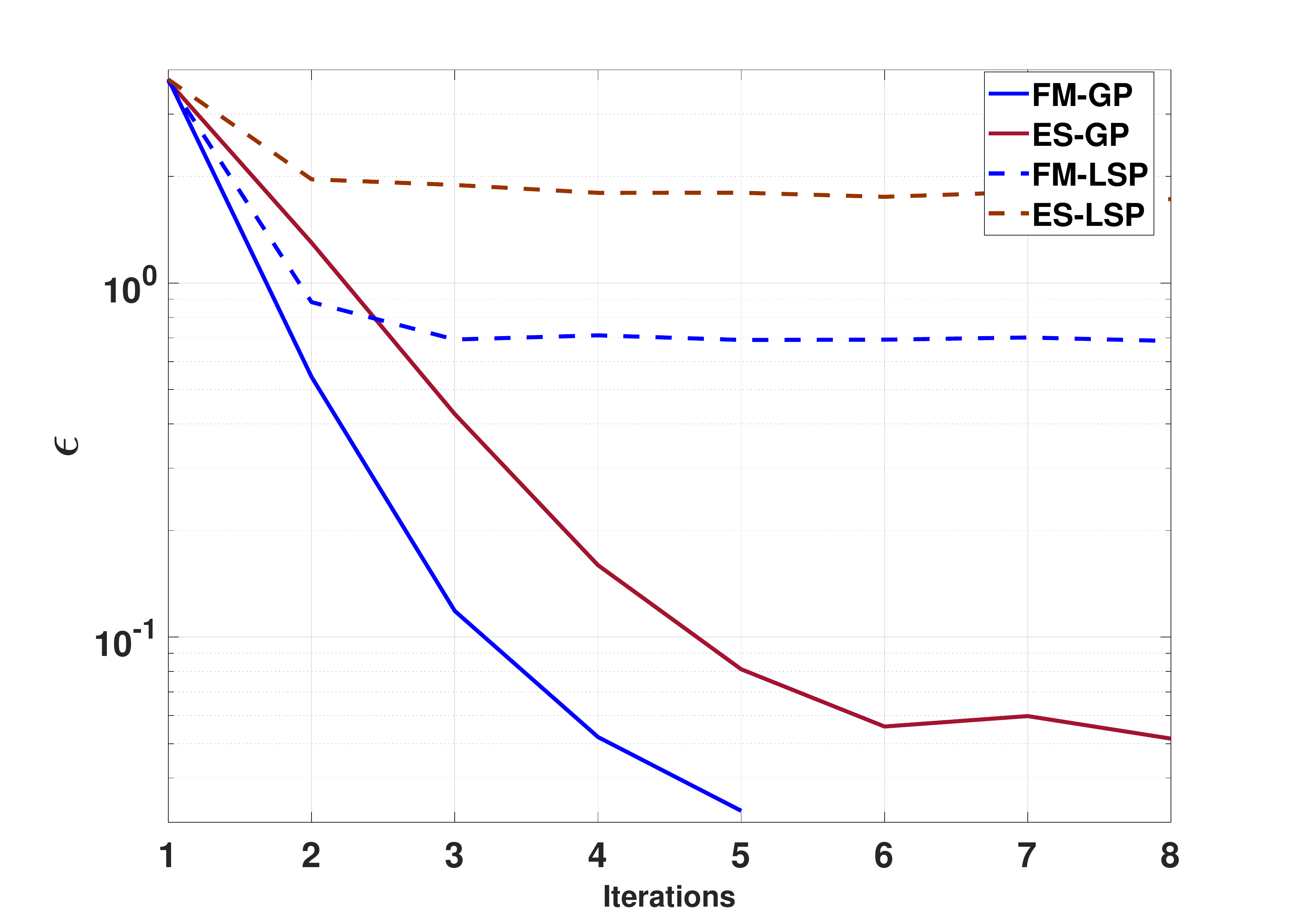}
    } \hskip -4.5ex
        \vspace*{-3mm}
    \caption{\small Convergence of the a) gradient projection (Algorithm 3) for a fixed ${\bf F}_{\bf BB}$, and b) hybrid precoding by alternating optimization to approximate the fully-digital precoder (Algorithm 2), $M_t = 128$, $M_r = 32$, $N_{RF} = 8$, $K = 4$.}
    \label{BP}
\end{figure}

Next, we show the beampattern generated at the BS corresponding to the full-digital precoders and the hybrid precoders in Fig.3. To demonstrate the phenomenon of beam misalignment between BS and RU beams, the beampatterns with the analog combiner (AC) for a RU generated for two cases, a) perfect alignment, and b) imperfect beam alignment with an alignment error of $1^o$, are shown. From Fig.3(a), the robust DP-FM and DP-ES are seen to have a uniform gain in the expected alignment error range and achieve a higher antenna gain in the case of beam misalignment compared to the non-robust CDP. A similar trend can be observed for the robust hybrid precoders FM-GP/LPS and ES-GP/LPS as shown in Fig.3(b).


\begin{figure}[!htb]
    \centering
   \subfigure[]
    {
            \includegraphics[scale=0.32]{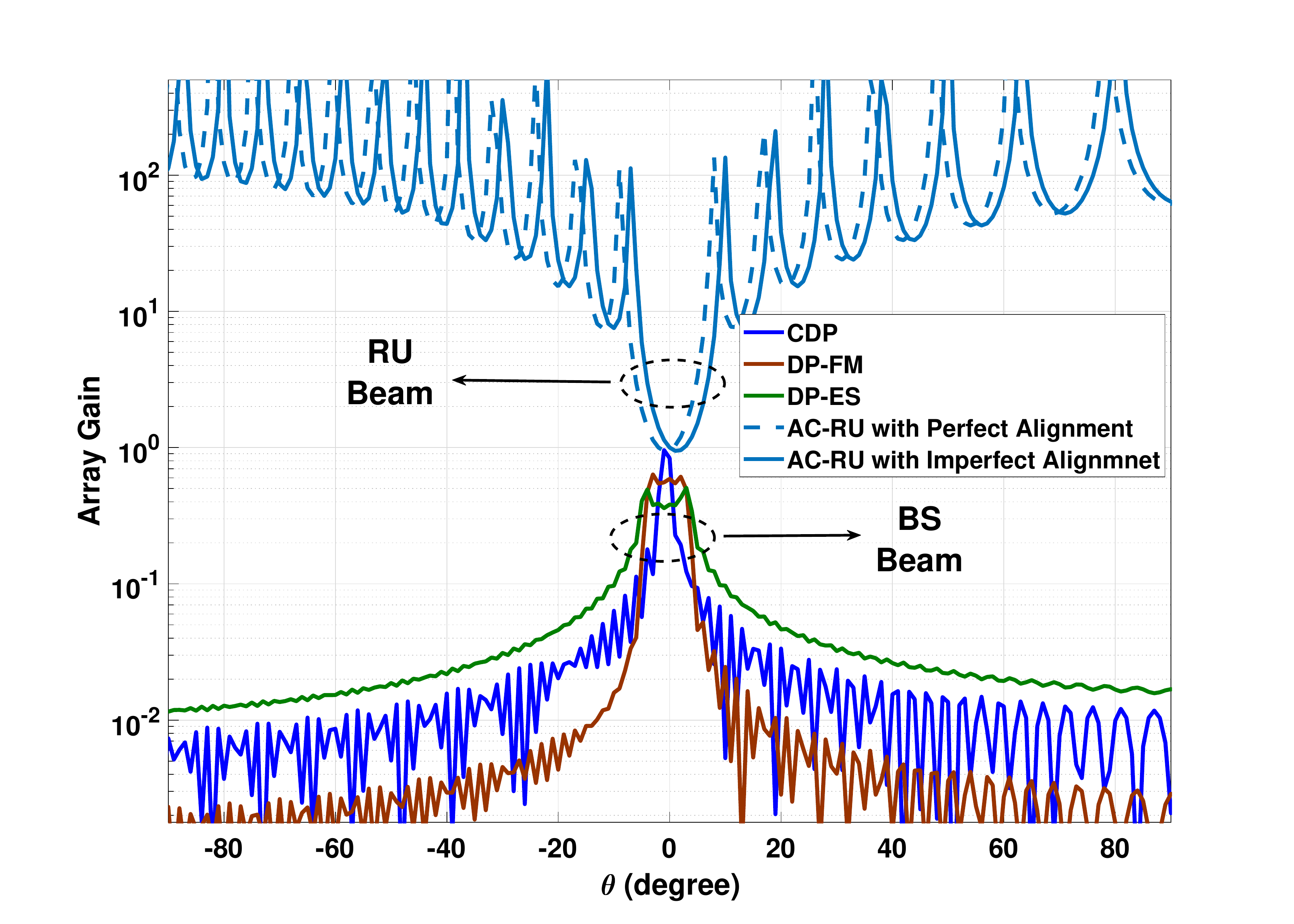}
    } \hskip -4.5ex
    \subfigure[]
    {
            \includegraphics[scale=0.32]{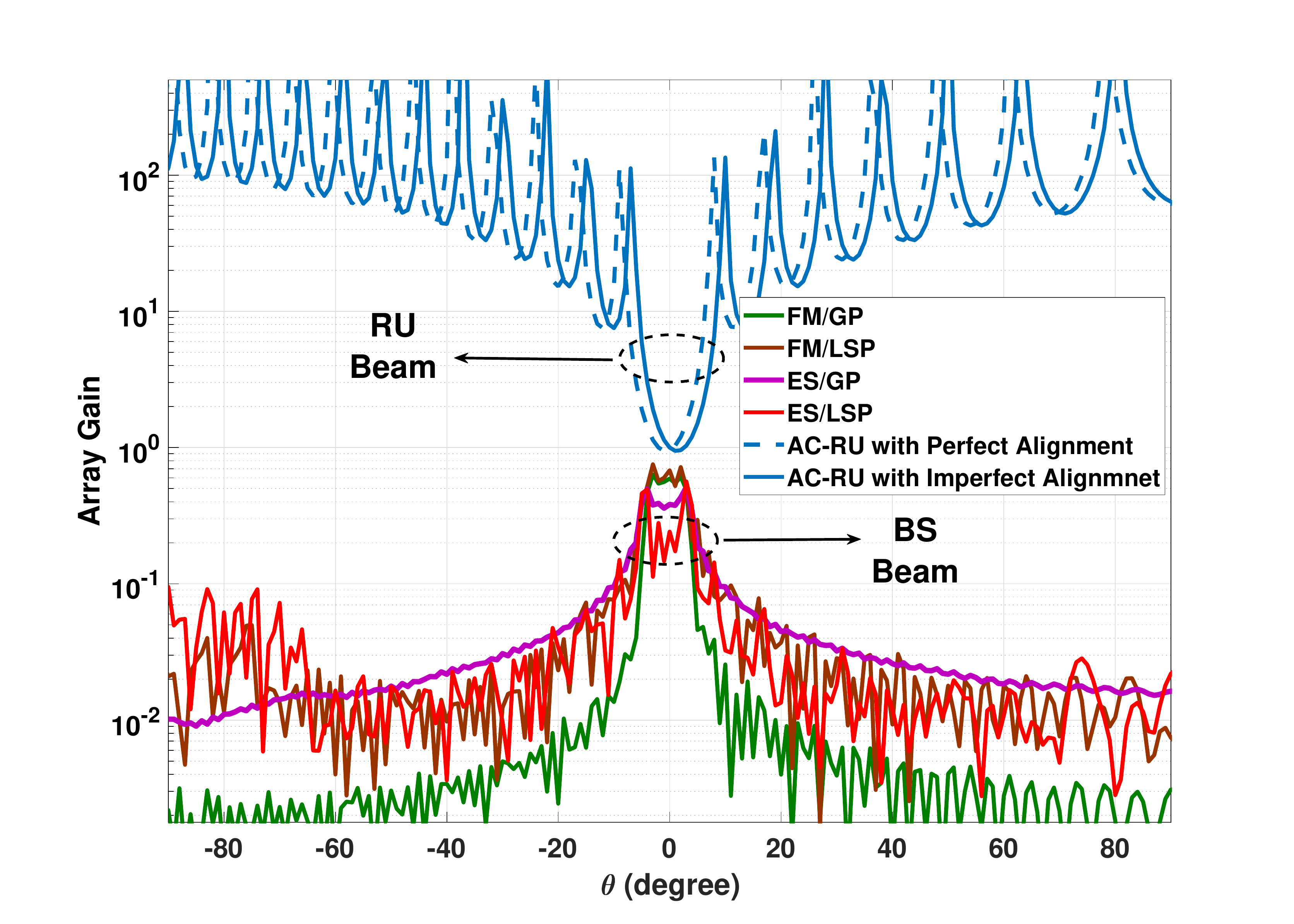}
    } 
        \vspace*{-3mm}
    \caption{\small Beam pattern at  the BS and RU for a) fully-digital precoders, and b) hybrid precoders approximating full-digital precoders $M_t = 128$, $M_r = 32$, $N_{RF} = 8$, $\delta = 1^o$.}
    \label{BP}
\end{figure}

    

In the following, we present the spectral efficiency performance. Given the received signal at the $k$-th RU, the  corresponding achievable rate  is obtained as 

\begin{equation}
R_k = \log_2 \left(1 + \frac{\frac{P}{K}|\mathbf{w}_k^H {\bf H}_k {\bf f}_k^{\bf HP}|^2}{\sigma_n^2 + \sum_{(j \neq k)} \frac{P}{K} {| {\bf w}_k^H {\bf H}_k {\bf f}_j^{\bf HP}|^2}} \right),
\end{equation}
and the overall spectral efficiency of the system is then given by $R_{SE} = \sum_{k=1}^K R_k$. Fig.4(a) and (b) present the performances of the hybrid precoding schemes. It can be observed that the robust hybrid precoders outperform the non-robust hybrid precoders in the presence of the beam misalignment. Specifically, a gain of at least $5 \; {\rm dB}$ and $3 \; {\rm dB}$ can be observed over the non-robust schemes for the hybrid precoding schemes based on `flat-mainlobe' and `error-statistics', respectively. While the performance difference between MO-based and GP-based schemes is marginal, the GP schemes require a much lower computational cost, as demonstrated in Table I. For the low-complexity FM-LSP and ES-LSP schemes, a performance loss of 0.6 dB and 2 dB can be observed compared to FM-GP and ES-GP schemes, respectively.


\begin{figure}[!htb]
    \centering
   \subfigure[]
    {
        \includegraphics[scale=0.32]{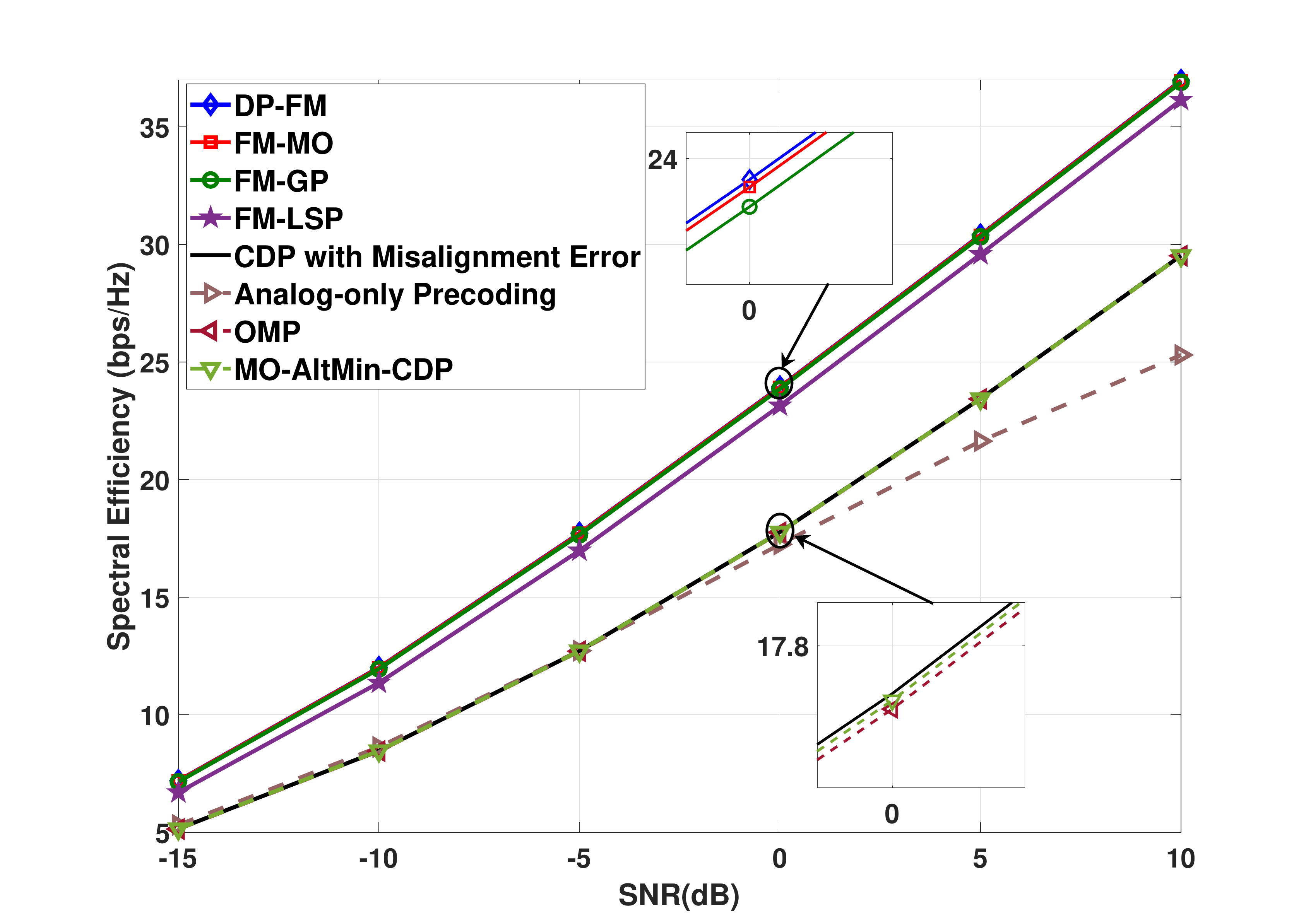}
    } \hskip -4.5ex
    \subfigure[]
    {
        \includegraphics[scale=0.32]{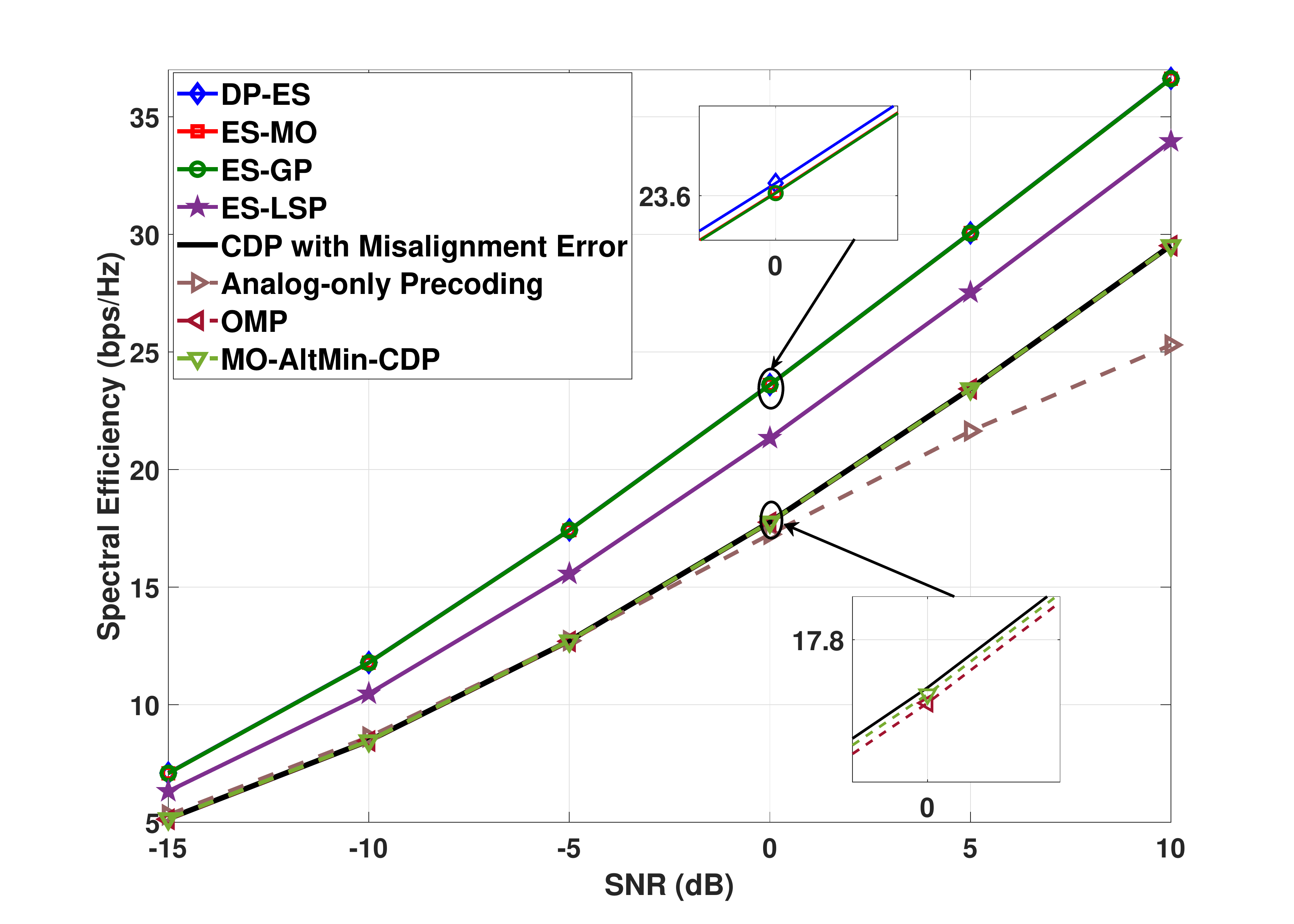}
    } \hskip -4.5ex
        \vspace*{-3mm}
    \caption{\small Spectral efficiency v.s. SNR for hybrid precoders approximating full-digital precoders based on a) `flat-mainlobe', and (b) `error-statistics', $M_t = 128$, $M_r = 32$, $N_{RF} = 8$, $K = 4$.}
    \label{BP}
\end{figure}

    

In Fig.5, we evaluate the performances of the proposed GP and LSP schemes with and without the second-stage digital precoder $\mathbf{F_{BD}}$. As illustrated in Fig.5, without $\mathbf{F_{BD}}$,  FM-GP/LSP and ES-GP/LSP schemes undergo a considerable performance loss due to the residual inter-receiver interference, especially at the higher SNRs. 


\begin{figure}[!t]
       \centering
       
               \includegraphics[scale=0.32]{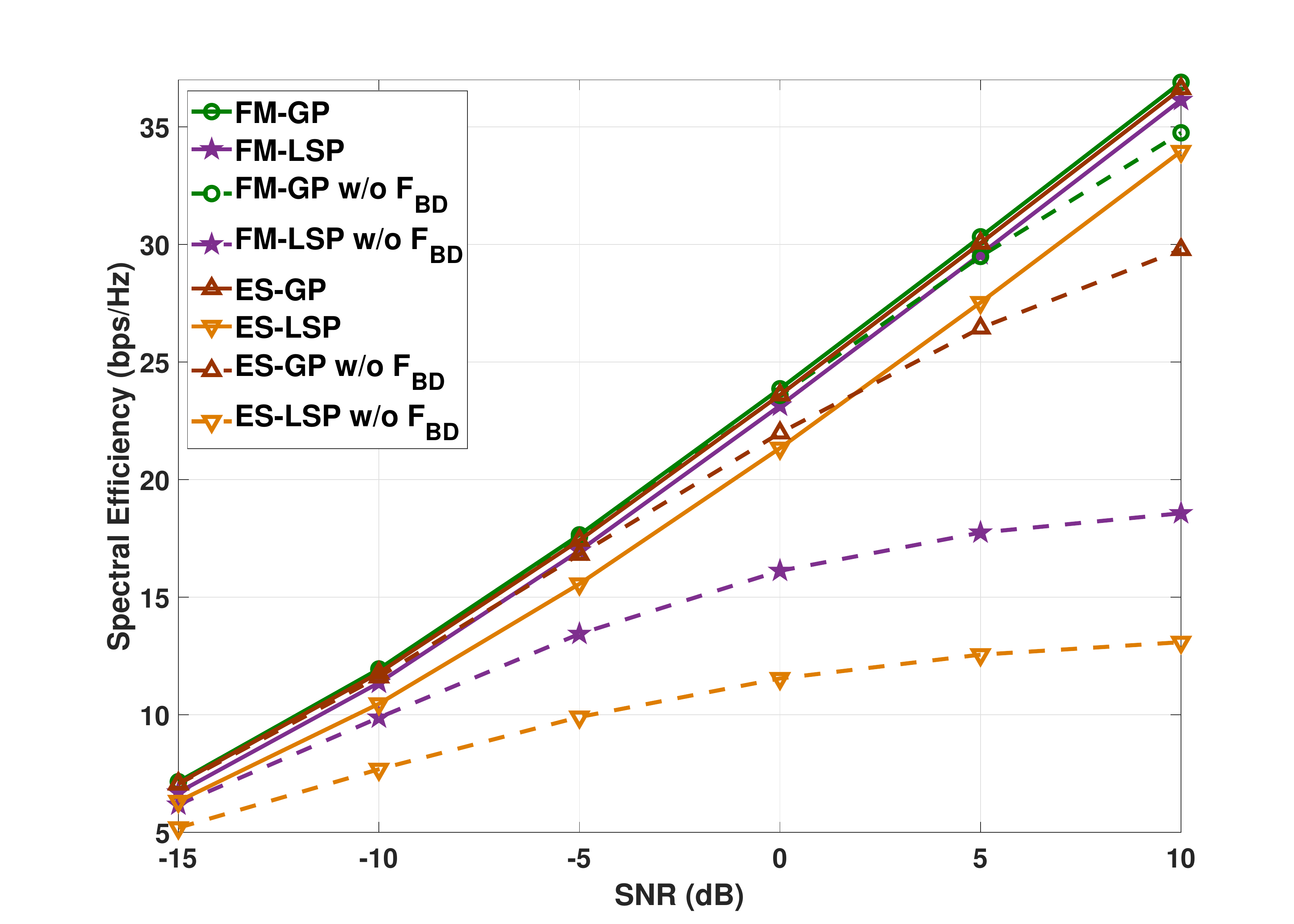}

\caption{\small  Spectral efficiency v.s. SNR with and without the second stage digital precoding for hybrid precoders approximating full-digital precoders, $M_t = 128$, $M_r = 32$, $N_{RF} = 8$, $K = 4$.}
    \label{SNR_LSP}
\end{figure}

\section{Conclusion}

In this paper, we have proposed robust hybrid precoding schemes to abate the performance loss owing to imperfect alignment between the beams at the transmitter and receiver. The proposed schemes, based on both the `flat-mainlobe' model and the `error statistics', involve an approximation of the robust full-digital precoder by the hybrid precoder. We used the GP algorithm to design the analog precoder and further proposed a low-complexity LSP method for the analog precoder design. A second-stage digital precoder is also introduced for managing the residual inter-receiver interference. Extensive numerical results validate the robustness of the proposed hybrid precoding schemes against imperfect beam alignment and the  performance  advantage  of  the  second-stage digital precoder. One of the future works can be the extension of the robust hybrid precoding design to multi-stream receivers for spectral and energy efficiency maximization.

\ifCLASSOPTIONcaptionsoff
  \newpage
\fi

\bibliography{myBib.bib}
\bibliographystyle{IEEEtran}

\end{document}